\documentclass[pra,twocolumn]{revtex4-2}
\usepackage{amsthm}
\usepackage{amsmath}
\usepackage{latexsym}
\usepackage{amsfonts}
\usepackage{amssymb}
\usepackage{color}
\usepackage{bbm,dsfont}
\usepackage{graphicx}
\usepackage{hyperref}
\usepackage{subfigure}
\usepackage{caption}
\usepackage{xr}
\usepackage{mathrsfs}
\usepackage[noend]{algpseudocode}
\usepackage{algorithmicx,algorithm}
\usepackage{graphicx}
\usepackage{[number}
\usepackage{float}
\usepackage{epstopdf}
\usepackage{mathrsfs}
\usepackage{appendix}

\definecolor{gray}{RGB}{123,123,123}
\UseRawInputEncoding

\newtheorem{proposition?}{Proposition?}

\theoremstyle{definition}

\newcommand{\tr}[1]{\textrm{tr}\left[#1\right]} 

\begin{document}
\title{Entanglement Verification with Deep Semi-supervised Machine Learning}
\author{Lifeng Zhang, Zhihua Chen}
\thanks{Electronic address:  chenzhihua77@sina.com}
\affiliation{School of Science, Jimei University, Xiamen 361021,China}
\author{Shao-Ming Fei}
\thanks{Electronic address:  feishm@cnu.edu.cn}
\affiliation{School of Mathematical Sciences, Capital Normal University, Beijing 100048, China,\\
Max Planck Institute for Mathematics in the Sciences, 04103 Leipzig, Germany}

\begin{abstract}
Quantum entanglement lies at the heart in quantum information processing tasks. Although many criteria have been proposed, efficient and scalable methods to detect the entanglement of generally given quantum states are still not available yet, particularly for high-dimensional and multipartite quantum systems. Based on FixMatch and Pseudo-Label method, we propose a deep semi-supervised learning model with a small portion of labeled data and a large portion of unlabeled data. The data augmentation strategies are applied in this model by using the convexity of separable states and performing local unitary operations on the training data. We verify that our model has good generalization ability and gives rise to better accuracies compared to traditional supervised learning models by detailed examples.
\end{abstract}

\maketitle

\section{Introduction}
Entanglement \cite{I1,I2} is an essential resource in many quantum information processing tasks such as quantum cryptography \cite{I3}, quantum communication \cite{I4}, measurement based quantum computation \cite{I5}, quantum simulation\cite{I6simu} and quantum metrology \cite{I7metro}. Detecting entanglement is an important and fundamental problem in the theory of quantum entanglement. Although many efforts have been made toward the detection of multipartite entanglement \cite{p1,RevMod,M6,M7,M8,M9,Gao,Hong1,Hong2,Hong3,fisher,Fei,M11,M12,M10,M13,M14,M15,p2,Entcer,Numwit,Mirwit}, the characterization of multipartite entanglement turns out to be still quite difficult both theoretically and experimentally \cite{Gurvits,Exp1,Exp2}. Appropriate entanglement witnesses \cite{M10,M13,M14,M15,p2,Entcer,Numwit,Mirwit} are usually constructed to detect entanglement in experiments. However, as the number of the particles increases, detecting multipartite entanglement becomes formidably difficult, even for special quantum states \cite{M16,M17}.

The machine learning aims to use mathematical or statistical models to obtain a general understanding of the training data to make predictions. It has been successfully applied  in quantum information theory theoretically and experimentally such as classification and quantification of quantum correlations, including quantum entanglement \cite{s1,s2, svm2,svm3,wei1,wei2,Gir}, EPR steerability \cite{s3} and quantum nonlocality \cite{s4,svm4,Dool,Poderini, Melnikov, Polino,Kriv}. However, most involved works adopt supervised training methods which requires a large number of labeled samples. The supervised methods inevitably fall into an intractable dilemma, i.e., the process of labeling a large number of quantum states is very time-consuming and almost impossible for multipartite and high-dimensional quantum correlations.

The semi-supervised machine learning was proposed to improve the classification performance by combining scarce labeled data with sufficient unlabeled data \cite{ss1}. Recently, the end-to-end semi-supervised model consisting of semi-supervised learning and deep learning modules was proposed, in which the network is trained by using the predictions on unlabeled data as pseudo-labels \cite{pseudo,mixmatch,fixmatch}. The safe semi-supervised support vector machine (S4VM) has been successfully used to detect steerability of two-qubit quantum states \cite{Lifeng}, though it is still time consuming to implement S4VM as the global simulated annealing search is needed.

In this paper, a deep semi-supervised learning model along with data augmentation strategies is proposed. Based on this model the bipartite entangled and separable states can be detected, the genuine tripartite entangled states, the tripartite entangled states and fully separable states can be classified simultaneously, and the white noise tolerances of $n$-qubit GHZ states in noise can be learned from the fuzzy bounds. Besides, the 2-separable  and genuine entangled states, 3-separable and 3-nonseparable states, 4-separable and 4-nonseparable states of the noisy $n$-qubit GHZ states are also classified via this model for $n=4,\cdots,10.$ The accuracy via our model is higher by using 30 (500) labeled states and 60 (10000) unlabeled states than the one via supervised learning by using 100 (4000) labeled states for the noisy $n$-qubit GHZ states (2-qubit states). The data augmentation strategies are performed by convex combination of the separable states, and by performing local unitary operations on the labeled states and unlabeled quantum states. Compared with the traditional supervised machine learning algorithms, the deep semi-supervised algorithm can achieve better accuracy and robustness. In addition, relatively high precision can be maintained with fewer labeled quantum states.

\section{Entanglement detection via deep semi-supervised learning method}
\subsection{Detection of $k$-nonseparability of multipartite quantum states}
Let $\mathcal{H}=\mathcal{H}_1\otimes\mathcal{H}_2\otimes\cdots\otimes\mathcal{H}_n$ be the Hilbert space of an $n$-partite system. For any partition $\{t_1,t_2,...,t_k\}$ of $\{1,...,n\}$, a quantum state $\rho\in\mathcal{H}$ is said to be $k$-separable if it can be expressed as a convex combination of the product states as follows,
\begin{eqnarray}\label{Eq1}
\label{rho}
&\rho=\sum\limits_{i}{p_{i}|{\psi^i_{t_1}}\rangle\langle\psi^i_{t_1}|
\otimes|{\psi^i_{t_2}}\rangle\langle\psi^i_{t_2}|\otimes \cdots \otimes|{\psi^i_{t_k}}\rangle\langle\psi^i_{t_k}|},
\end{eqnarray}
where $\sum\limits_{i=0}{p_{i}}=1$ and $p_{i}$ is a classical probability distribution, $|{\psi^i_{t_j}}\rangle$ is a pure state in the subsystem $t_j$.
Otherwise $\rho$ is said to be $k$-nonseparable.

The $k$-separability of a state $\rho$ can be detected by entanglement witness \cite{p1}. A state $\rho$ is $k$-nonseparable if there is a hermitian operator $W$ satisfying $tr(\tau W)\leq 0$ for all $k$-separable states $\tau$, but $tr(\rho W)>0$ \cite{p2}. Nevertheless, finding a suitable witness for arbitrary high-dimensional quantum states is extremely complex and impractical. Instead of entanglement witness, inequalities involving some entries of a density matrix, criteria based on local uncertainty relations and fisher information were also used to detect the $k$-nonseparability \cite{Gao,Hong1,Hong2,Hong3}, although they are far from being perfect because of the complex structure of multipartite entanglement. We propose below a semi-supervised deep neural network method to detect the bipartite and multipartite entanglement of quantum states.

\subsection{The deep semi-supervised neural networks}
Based on Pseudo-Label \cite{pseudo} and FixMatch \cite{fixmatch}, we present the deep semi-supervised neural network approach which can be used in multi-classification problems with a small number of labeled samples and a large number of unlabeled samples.

\textbf{\textit{Consistency Regularization}} Consistency regularization is a commonly used technique for semi-supervised learning, which can force the model to become more confident in predicting labels on unlabeled data, by encouraging the model to produce the same output distribution when the input or weights are slightly perturbed \cite{con1,con2,con3}. Data augmentation can be applied to semi-supervised learning based on consistency regularization.

\textbf{\textit{Augmentation Strategy}} Augmentation methods are used for quantum states inspired by augmentation strategies such as flipping and shifting in image processing.
Two augmentation strategies are applied based on two properties of quantum states: invariance under local unitary operations for both entangled and separable quantum states,
and the convexity of the set of separable states.

Under local random  unitary operations, a quantum state $\rho$ becomes
\begin{eqnarray}
\label{lu}
&\rho^\prime=(V_1\otimes \cdots \otimes V_n)\rho(V_1\otimes \cdots \otimes V_n)^{+},
\end{eqnarray}
where $V_i$ $(i=1 \cdots n )$ are arbitrary random unitary matrices on the $i-\rm{th}$ system, and $``+"$ represents conjugation and transformation. Let $\mathcal A_k(\rho)$ be the quantum state after performing the $k^{\rm{th}}$ local random unitary transformation on $\rho$. Specifically, $\mathcal A_0(\rho)=\rho$. Let $\mathcal{X}$ be a set of quantum states and $\mathcal M(\mathcal{X})$ be a random convex combination of all separable states in $\mathcal{X}$, which can generate an equal number of new separable states as $\mathcal{X}$. It is worth noting that our augmentation methods are all lossless.


Denote the labeled data as $\mathcal{X}$, i.e. $\mathcal{X}=\{x_i, y_i\}_{i=1}^l$ and  the unlabeled data as $\mathcal{U}$, i.e. $\mathcal{U}=\{x_j\}_{j=l+1}^{l+u}$, where $y_i$ is the one-hot label. Applying the augmentation technique by local unitary operations, we attain $K$ augmented labeled data sets $\mathcal{X}^\prime=\{x_{i,k}, y_{i,k}\}_{i=1,\cdots,l}^{k=0,\cdots,K}$ with $x_{i,k}=\mathcal A_k(x_i)$ and $y_{i,k}=y_{i}$, unlabeled data $\mathcal{U}^\prime=\{\hat{x}_{j,k}\}_{j=l+1,...l+u}^{k=0,...K}$ with $\hat{x}_{j,k}=\mathcal A_k(x_j)$.

\textbf{\textit{Guess-Label}} Guess-labels of the samples in $\mathcal{U}^\prime$ are attained via the model's predictions, which are assumed to be the target classes of the samples and will be used to compute the unsupervised loss term.

Firstly the averaged prediction $y^{\prime}_{j}$ of $x_j$ can be attained,
\begin{eqnarray}
\label{p1}
&y^{\prime}_{j}=\frac{1}{K+1}\sum\limits_{k=0}^K{p_{model}(y|\mathcal A_k(x_j))},~~~ x_j\in \mathcal{U},
\end{eqnarray}
where $p_{model}(y|x)$ refers to a generic model which produces a distribution over the class labels $y$ for an input $x$.

Then the guess-label $\hat{y}_{j}$ can be attained as follows,
\begin{eqnarray}
\label{p2}
&\hat{y}_{j}=argmax\{(y^{\prime}_{j}>\tau)y^{\prime}_{j}\},
\end{eqnarray}
where $\tau$ is the threshold. The sample is retained when the average prediction of a sample is greater than $\tau$. $\hat{y}_{j}$ can be regarded as the real label. Then we have the new unlabeled data $\hat{\mathcal{U}}=\{\hat{x}_{j,k}, \hat{y}_{j,k}\}_{j\in\{l+1,\cdots,l+u\}}^{k\in\{0,\cdots,K\}}$ with $\hat{y}_{j,k}=\hat{y}_j$. $\mathcal{X}^\prime$ and $\hat{\mathcal{U}}$ are used to compute the labeled and unlabeled loss terms, respectively, during the network training process. Furthermore, a valid ``one-hot" probability distribution can be generated by applying the argmax function in Eq. \eqref{p2}, which makes guess-labels be very close to the entropy minimization and the model's low-entropy predictions on the unlabeled data \cite{em,em1}.

\textbf{\textit{Loss Function}} The loss function $\mathcal L$ consists of two cross-entropy loss terms: a supervised loss $\mathcal L_s$ and an unsupervised loss $\mathcal L_u$,
\begin{eqnarray}
\label{loss}
&\mathcal L=\mathcal L_s+\lambda_u\mathcal L_u,\\ \nonumber
&\mathcal L_s=\frac{1}{|\mathcal X^{\prime}|}\sum_{i=1}^{|\mathcal X^{\prime}|}H(y_{i,k}, p_{model}(y|x_{i,k})),\\ \nonumber
&\mathcal L_u=\frac{1}{|\hat{\mathcal{U}}|}\sum_{j=1}^{|\hat{\mathcal{U}}|}H(\hat{y}_{j,k}, p_{model}(y|x_{j,k}))\\ \nonumber
\end{eqnarray}
with $i=1,...,l$, $j\in(l+1,...,l+u)$ and $k\in(0,...,K)$, where $\lambda_u$ is a hyper-parameter representing the relative weight of the unlabeled loss, $H(p,q)$ is the standard cross-entropy between two probability distribution $p$ and $q$.
The  entire algorithm is summarized in Algorithm~\ref{algorithm1} and the codes are listed in\cite{code}:
\begin{algorithm}[H]
\caption{The $t^{\rm{th}}$ semi-supervised model process} 
\label{algorithm1}
 {\bf Input:} 
 {Labeled data $\mathcal{X}=\{x_i,y_i\}_{i=1}^l,$ unlabeled data $\mathcal{U}=\{x_j\}_{j=l+1}^{l+u},$ unlabeled loss weight $\lambda_u$, confidence threshold $\tau$, the number of augmentations K. } \\
 {\bf Output:} 
{$loss$ $L$}
\begin{algorithmic}[1]
\For{$i=1$ to $l$}
   \For{$k=0$ to $K$}
      \State $x_{i,k}=\mathcal A_k(x_i)$
      \State $y_{i,k}=y_i$
   \EndFor  ¡¡¡¡
\EndFor
\State ${\mathcal X}^{\prime}=\{x_{i,k}, y_{i,k}\}_{i=1,\cdots,l}^{k=0,\cdots,K}.$
\For{$j=l$ to $l+u$}
   \For{$k=0$ to $K$}
      \State $\hat{x}_{j,k}=\mathcal A_k(x_j)$
   \EndFor
\EndFor
\State ${\mathcal U}^{\prime}=\{\hat{x}_{j,k}\}_{j=1,\cdots,U}^{k=0,\cdots,K}$
\For{$j=l$ to $l+u$}
   \State $y^{\prime}_{j,k}=\frac{1}{K+1}\sum_{k=0}^{K}{p_{model,t-1}(y|\hat{x}_{j,k})}$ {\textcolor{gray}{// $p_{model,0}$ is the supervised model trained with $\mathcal{X}$}}
   \State $\hat{y}_{j,k}=argmax\{(max(y^{\prime}_{j,k})>\tau)y^{\prime}_{j,k}\}$
\EndFor
\State $\hat{\mathcal{U}}=\{\hat{x}_{j,k}, \hat{y}_{j,k}\}_{j\in\{l+1,\cdots,l+u\}}^{k\in\{0,\cdots,K\}}$
\State $L_s=\frac{1}{|\mathcal X^{\prime}|}\sum H(y_{i,k}, p_{model,t}(y|x_{i,k}))$ {\textcolor{gray}{// Cross-entropy loss for $\mathcal X^{\prime}$}}
\State $L_u=\frac{1}{|\hat{\mathcal{U}}|}\sum H(\hat{y}_{j,k}, p_{model,t}(y|\hat{x}_{j,k}))$ {\textcolor{gray}{// Cross-entropy loss for $\hat{\mathcal{U}}$ with high-confidence guess-labels }}
\State \Return $L=L_s+\lambda_u L_u$
\end{algorithmic}
\end{algorithm}

\begin{figure}[H]
\centering
\includegraphics[width=8cm]{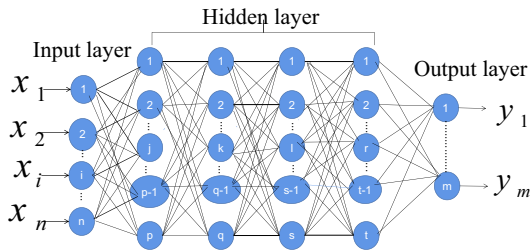}
\caption{Neural network structure consists of an input layer, an output layer and four hidden layers. The activation of the output layer is softmax function, and the activations of  other layers are all ReLu function.}
\label{fig0}
\end{figure}

\textbf{\textit{Network Architecture}} We build two neural network architectures with different structures for two qubit and noisy three qubit GHZ states, respectively, as shown in FIG. \ref{fig0}. From the input layer to the output layer of the network, the number of the nodes are 16, 512, 256, 128, 16, 2 (64, 512 256,128,16,3) respectively for two-qubit (noisy 3-qubit GHZ states) entanglement detection.

\textbf{\textit{Hyperparameters}} Concerning the entanglement detection problem, the training epochs of the neural network are shown in the appendix. The maximum number of updates $T$ is 30, and  $\tau\geq 0.9$. Furthermore, the unlabeled loss weight $\lambda_u$ slowly increases and then decreases from a starting point 0.15 when updating the model each time. Through the process the local minima can be effectively attained and the pseudo-labels of unlabeled data can be as close to the real labels as possible without using the optimization process \cite{pseudo}.

\section{Neural Network Training}
\subsection{Preparation of training samples}
For two-qubit states, the density matrices are generated as follows:  two random 4$\times$4 matrices $N_1$ and $N_2$ are generated, which are then used to generate a Hermitian matrix $H=(N_1+iN_2)(N_1+iN_2)^+$, where $``+"$ stands for conjugate and transpose. Then a density matrix is given by $\rho=\frac{H}{\tr H}$.

The training samples are generated in the following steps:

{\bf{1)}} Firstly $l_1$ random density matrices $\rho$ can be created. One-hot labels can be attained via the PPT criterion \cite{I2}. If $\rho$ is separable, $y=[1,0]$, otherwise $y=[0,1]$. Obviously, we can collect a batch of labeled data $\mathcal{L}=\{\rho_i ,y_i\}_{i=1}^{l_1}$ with  half being entangled states and half being separable states.

{\bf{2)}} Secondly $u_1$ random density matrices are generated as the unlabelled states, $\mathcal{U}=\{\rho_j\}_{j=l_1+1}^{l_1+u_1}$. As the numbers of entangled and separable states are in imbalance, there is a common problem of uneven distribution of category data on  $\mathcal{U}$. For the sake of balance, some random separable states are generated and the strategies $\mathcal A$ and $\mathcal M$ are used to increase the separable states in $\mathcal{U}$.

The n-qubit GHZ state mixed with white noise is the form,
\begin{eqnarray}
\label{ghz}
&\rho_{ng}=p|\text{GHZ}_n\rangle\langle\text{GHZ}_n|+\frac{1-p}{2^n}\mathcal{I},
\end{eqnarray}
where $|\text{GHZ}_n\rangle=\frac{1}{\sqrt{2}}({|0\rangle}^{\otimes n}+{|1\rangle}^{\otimes n})$ and $\mathcal{I}$ is the $2^n\times 2^n$ identity matrix. $\rho_{ng}$ is  biseparable if and only if $p\le b_2$ \cite{M7} and fully-separable  if and only if $p\le b_n$ \cite{sep,sep1} with $b_2=\frac{2^{n-1}-1}{2^n-1}$ and $b_n=\frac{1}{1+2^{n-1}}$. $\rho_{ng}$ is $k$-separable if only if $p\le b_k$ with $b_k=\frac{1}{1+\frac{2k-n}{n}2^{n-1}}$ when $k\geq\frac{n+1}{2}$ \cite{p2}.

For this noisy n-qubit GHZ state, the training samples are generated by the following steps:

{\bf{i)}} For each $ n\geq 3$, the labeled set $\mathcal{V}=\{\rho_{ng}^i ,y_i\}_{i=1}^{l_2}$ is generated randomly and we attain equal number of k-separable states. Here $y_i$ is a $n-$ dimensional vector, if $\rho_{ng}^i$ is $k$-separable, $y_i=[0,...,0,1,0...,0]$ with the $k^{th}$ entry  1 and  other entries  $0$, e.g., $y_i=[0,0,...,0,1]$ ($y=[0,1,0,...,0]$) if $\rho_{ng}^i$ is fully separable(biseparable).

{\bf{ii)}} The unlabeled set $\mathcal{W}=\{\rho_{ng}^j\}_{j=l_2+1}^{l_2+u_2}$ is generated randomly for $0\le p\le1$, together with some more fully entangle states and $k$-separable states generated randomly. Then we perform augmentation strategies $\mathcal A$ and $\mathcal M$ to increase the number of the $k$-separable$(k=2,...,n)$ states in order to balance unlabeled data $\mathcal{W}$.

\subsection{Training the neural networks via semi-supervised learning.}
Taking the entanglement detection of two-qubit quantum states as the example, the model training and  updating process can be divided into the following steps:

{\bf{a)}} For labeled data $\mathcal{L}$ and unlabeled data $\mathcal{U}$, we generate $K$ augmentations as $\mathcal{L}^\prime=\{\rho_{i,k},y_{i,k}\}_{i=1,\cdots,l_1+1}^ {k=0,\cdots,K}$ and $\mathcal{U}^\prime=\{\rho_{j,k}\}_{j=l_1,\cdots,l_1+u_1}^{k=0,1,\cdots,K}$ with $\rho_{i,k}=\mathcal{A}_k(\rho_i),$  $\hat{y}_{i,k}=y_i$ for $\mathcal{L}$ and $\rho_{j,k}=\mathcal{A}_k(\rho_j)$ for $\mathcal{U}$. The quantum state $\rho_i$ can be expressed by the feature vectors $F:$ $x_i=\{\text{Tr}\left[(\sigma_k\otimes \sigma_l)\rho_i\right]\}_{k,l=0}^3,$ where $\sigma_{k}$ and $\sigma_l$ are the standard Pauli matrices and $\sigma_0$ is $2\times 2$ identity matrix. To improve the efficiency of entanglement detection for arbitrary 2-qubit quantum states with the help of machine learning, we use the partial information $F_1:$ $\{\langle A_i\otimes B_j\rangle\}_{i,j=1,2,3}$ with $A_i=\sigma_i$ and $B_j=\sigma_j$, or the partial information  $F_2:$ $\{\langle A_i\otimes B_j\rangle\}_{i=1,2,3}^{j=1,2}$ with $A_i=\sigma_i,$ $B_1=\frac{\sigma_1+\sigma_3}{\sqrt{2}}$ and $B_2=\frac{\sigma_1-\sigma_3}{\sqrt{2}}$
as the feature vectors.

{\bf{b)}} Predictions of the unlabeled data $\hat{\mathcal{U}}$ can be attained by supervised learning by using the labeled data $\hat{\mathcal{L}}$ as the input to the network, as shown in FIG. \ref{fig0}. Firstly the network shown in FIG. \ref{fig0} is trained by minimizing Eq. \eqref{loss} ($\lambda_u=0$) via Adam optimizer in python with the fixed epochs of $\hat{\mathcal{L}}$ and a fixed learning rate $I_r=0.003$. Then the average predictions are attained by Eq. \eqref{p1}. Lastly, we calculate the guess-labels of samples with the average predictions greater than the threshold $\tau$ by Eq. \eqref{p2}. Thus a subset $\hat{\mathcal{U}}^\prime$ of the set of $\hat{\mathcal{U}}$ with pseudo-labels are attained. $\hat{\mathcal{U}}^\prime$ is used to calculate the loss terms of unlabeled data in Eq. \eqref{loss} at the next step of the model training.

{\bf{c)}} Calculating new $\hat{\mathcal{U}}^\prime$ with pseudo-labels through updating semi-supervised model $T$ times. By training the semi-supervised model using  $\hat{\mathcal{U}}^\prime$ $(t-1){\rm{th}}$ time ($\hat{\mathcal{U}}^\prime$ from {\bf{b)}} when $t=1$) as the input to the network shown in FIG.\ref{fig0}, the new predictions of the unlabeled data $\hat{\mathcal{U}}$ for the $t\rm{th}$ time can be attained. Here, $\lambda_{u}=\frac{e^{-s_t^2}}{2\pi}$, $s_t=t\frac{\|a-b\|}{T}$ $(t=1,...,T)$ with $a=-1$, $b=1$ and $T$ the total number of updates for the semi-supervised model. With the same process  as {\bf{b)}}, firstly the network shown in Fig.\ref{fig0} is trained by minimizing Eq. \eqref{loss} via Adam optimizer in python with the fixed epochs of $\hat{\mathcal{L}}$ and $\hat{\mathcal{U}}^\prime$, and fixed learning rate $I_r=0.003$. Then the new average predictions are attained by Eq. \eqref{p1}. Lastly a new $\hat{\mathcal{U}}^\prime$ can be attained by Eq. \eqref{p2}. The best model is selected by the validation set to verify the performance of these models.

\section{Numerical Results}
\subsection{Detecting entanglement of two-qubit quantum states}
\begin{figure}[H]
\centering
\includegraphics[width=7.5cm]{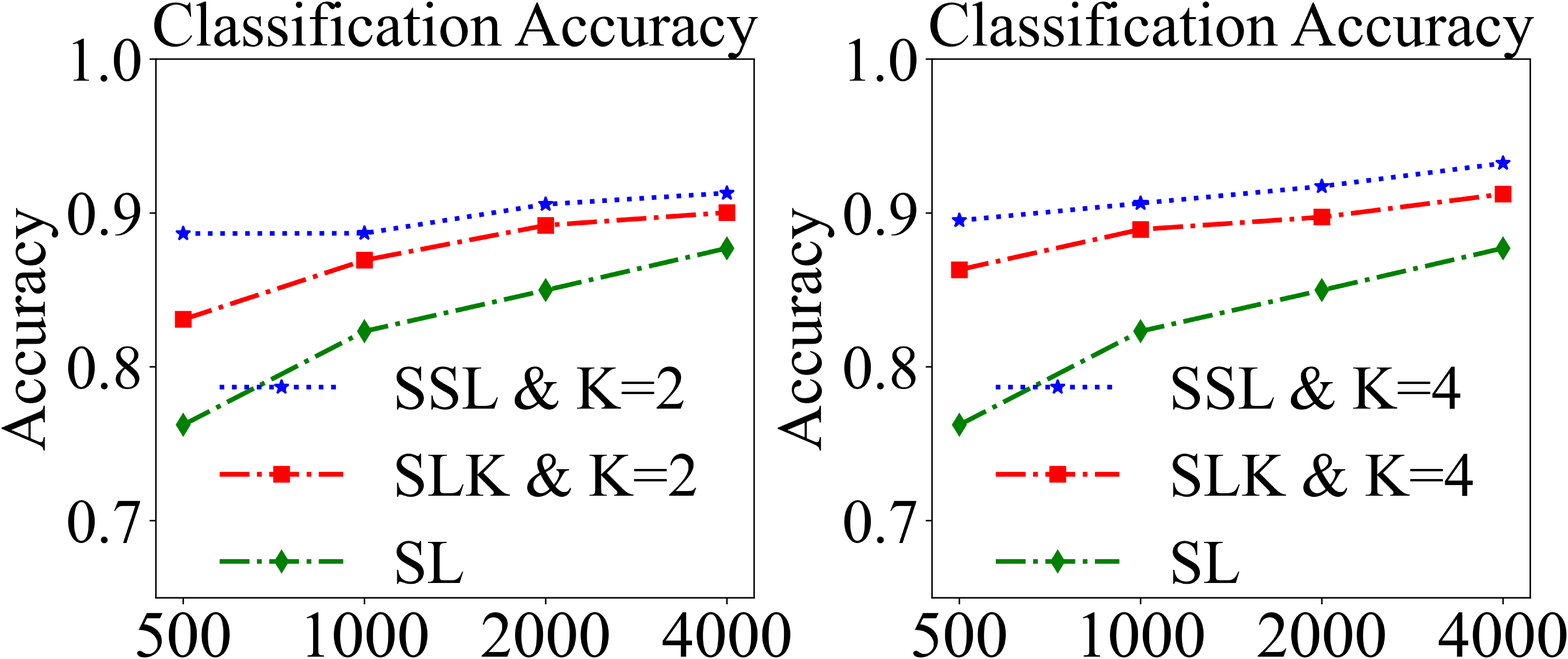}
\caption{Classification accuracy from SSL, SLK and SL by using  $l$ labeled and $u$ unlabeled quantum states
are represented by the dotted purple line with $\star$, orange dash-dot line with $\Box$ and green dash-dot line with $\diamond$, respectively, when $K=2$ and $4,$ $l=500,1000,2000$ and $4000$ and $u=10000,20000,40000$ and $80000$.}
\label{2-accuracy}
\end{figure}

\begin{figure}[H]
\centering
\includegraphics[width=7.5cm]{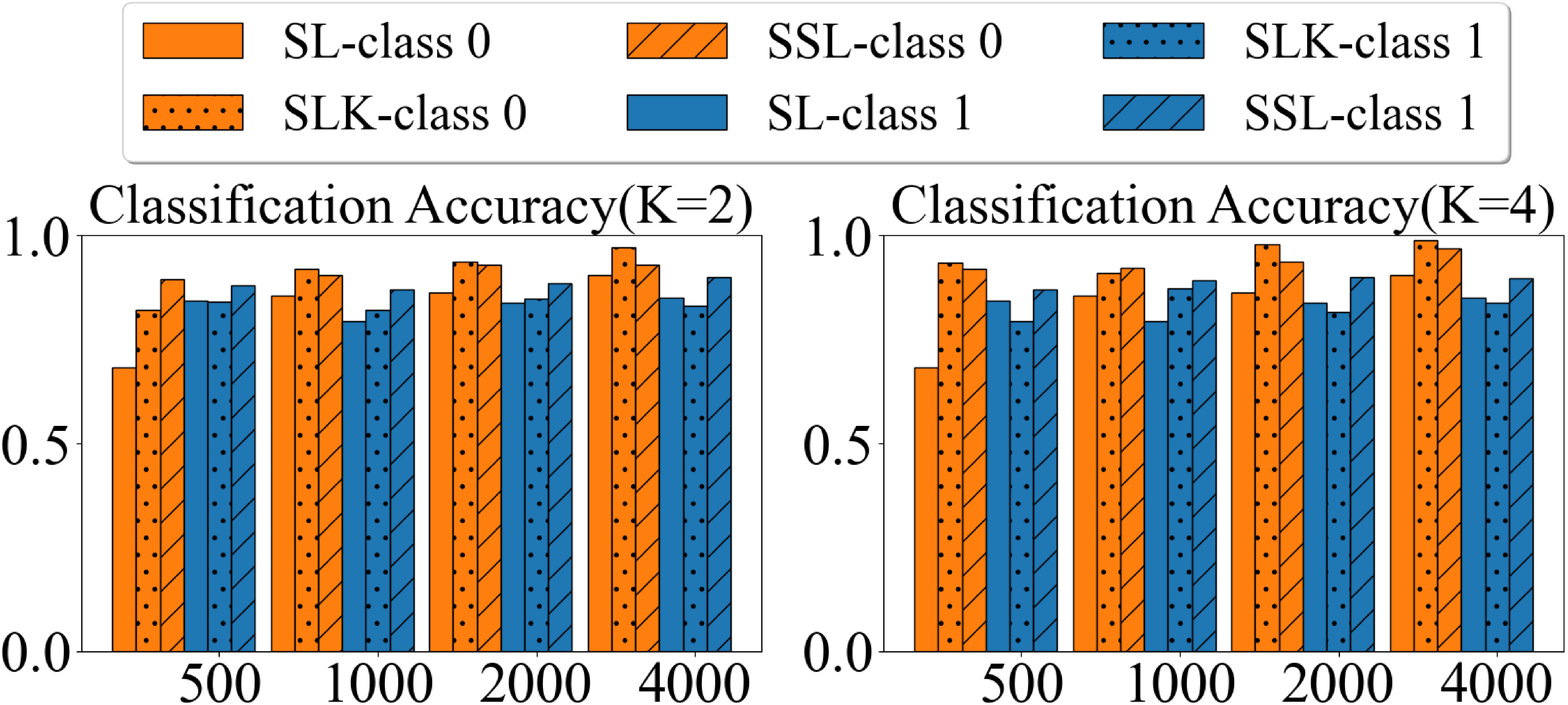}
\caption{Classification accuracy from SSL, SLK and SL  by using $l$ labeled and $u$ unlabeled quantum states when $K=2$ and $4,$  $l=500,1000,2000$ and $4000$ and $u=10000,20000,40000$ and $80000$. Each set of three orange columns represents the classification accuracy on 3000 separable states (class-0) from SL, SLK and SSL, respectivly; each set of three blue columns represents the classification accuracy on 3000 entangled states (class-1) from SL, SLK and SSL, respectively.}
\label{2-classa}
\end{figure}

\begin{figure}[H]
\includegraphics[width=8.5cm,]{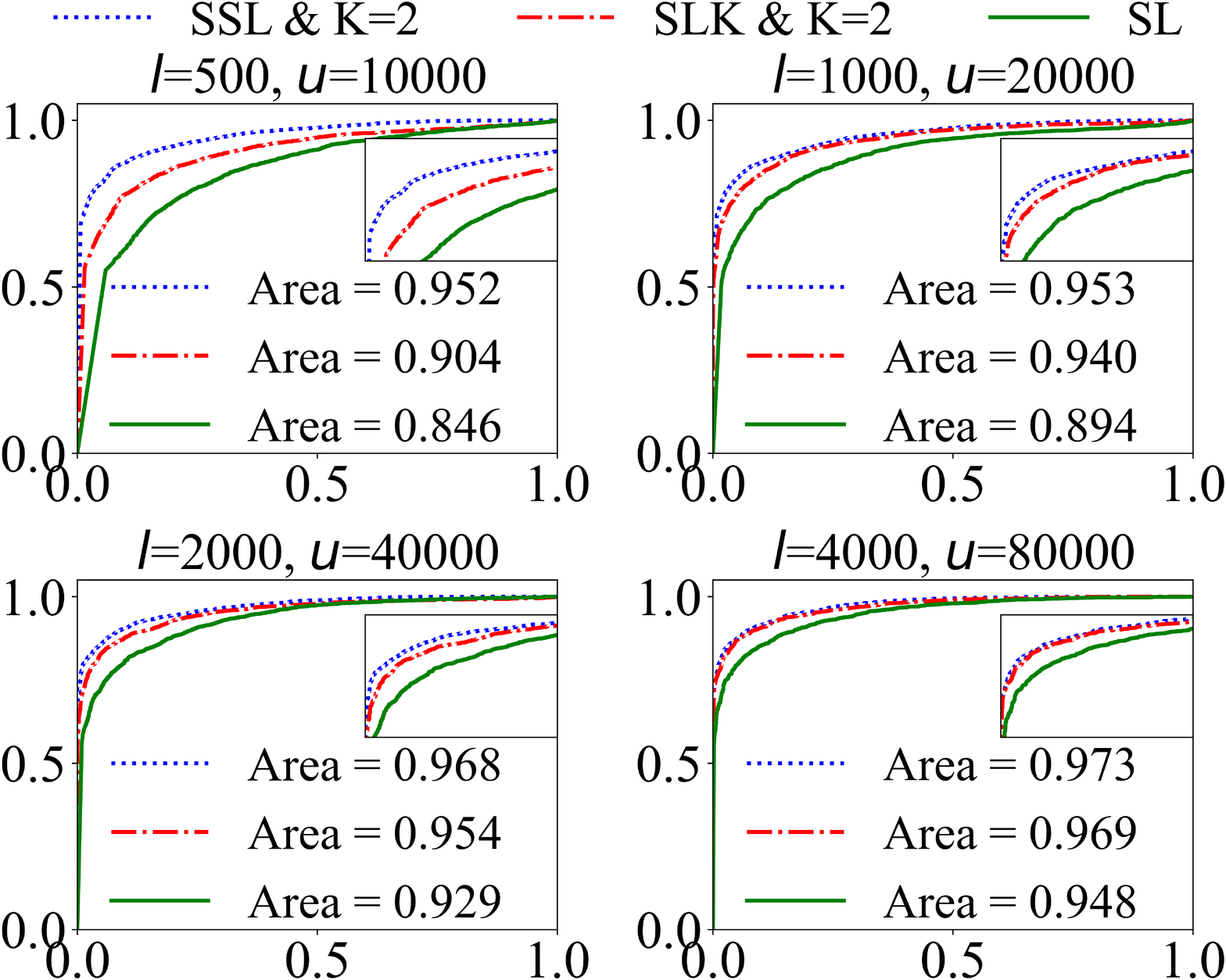}
\caption{The ROC curves of the test samples by SSL, SLK and SL are represented by the blue dotted line, red dash-dotted line and green solid line, respectively, for $l=500,1000,2000$ and $4000,$ $u=10000,20000,40000$ and $80000$ and $K=2$,the horizontal axis represents the false positive rate and the longitudinal axis represents the positive rate.}
\label{2-roc-k2}
\end{figure}

\begin{figure}[H]
\centering
\includegraphics[width=8.5cm,]{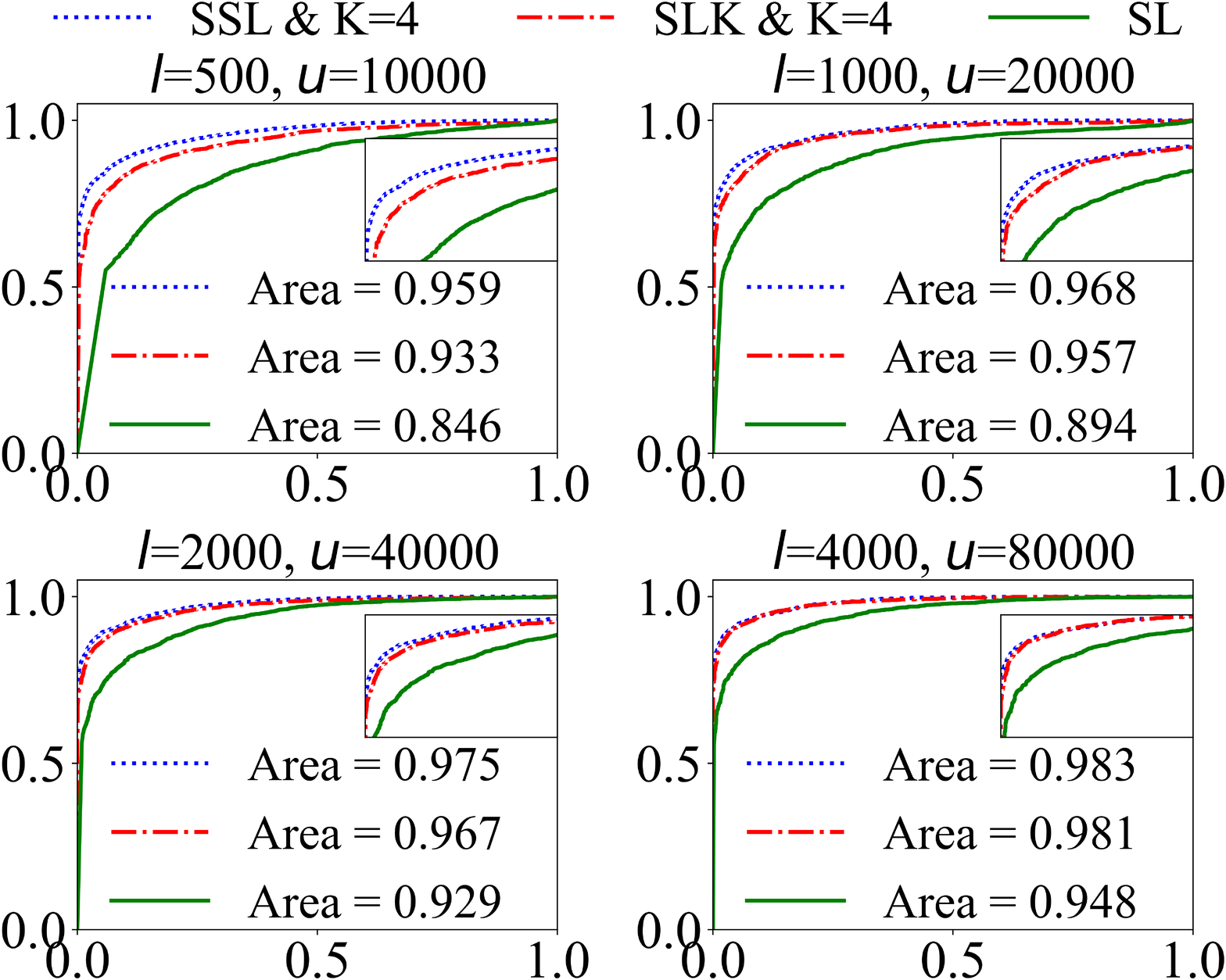}
\caption{The ROC curves of the test samples by SSL, SLK and SL are represented by the blue dotted line, red dash-dotted line and green solid line, respectively, for $l=500,1000,2000$ and $4000,$ $u=10000,20000,40000$ and $80000$ and $K=4$, the horizontal axis represents the false positive rate and the longitudinal axis represents the positive rate.}
\label{2-roc-k4}
\end{figure}

For two-qubit quantum states, the classification accuracy on the test samples consisting of 3000 random separable states and 3000 random entangled states are shown in
FIG.\ref{2-accuracy}, \ref{2-classa}, \ref{2-roc-k2} and \ref{2-roc-k4}.

In order to show the performance of our algorithm, we compare the results obtained by our algorithm (semi-supervised learning (SSL)), supervised learning with $K$ augmentations on labeled samples (SLK) and supervised learning without augmentations (SL). Let $l=500,1000, 2000$ and $4000$, and $u=20l$. The classification accuracy is shown in the FIG. \ref{2-accuracy} and FIG.\ref{2-classa}.
\begin{table}[H]
\centering
\caption{The difference between the accuracy on the test samples (separable states, entangled states) from SSL or SLK and SL by using $l$ labeled quantum states and $u$ unlabeled quantum states with $K=2,$ represented by
$\Delta_{SSL}$ $(\Delta_{SSL}^{-}, \Delta_{SSL}^{+})$ and $\Delta_{SLK}$ $(\Delta_{SLK}^-,\Delta_{SLK}^+)$.}
\label{tab1}
\begin{tabular}{|l|l|l|l|l|l|l|l|l|}
\hline
$l$  &u&$\Delta^{-}_{SSL}$&$\Delta^{-}_{SLK}$&$\Delta^{+}_{SSL}$&$\Delta^{+}_{SLK}$&$\Delta_{SSL}$&$\Delta_{SLK}$ \\ \hline
500  &10000 & 21.27\% & 13.83\%  & 3.6\%    & -0.13\% & 12.43\%  &   6.85\% \\ \hline
1000 &20000 & 4.97\%  & 6.5\%    & 7.77\%   & 2.73\%  & 6.37\%   &   4.62\% \\ \hline
2000 &40000 & 6.53\%  & 7.4\%    & 4.63\%   & 1.03\%  & 5.58\%   &   4.22\% \\ \hline
4000 &80000 & 2.33\%  & 6.63\%   & 4.83\%   & -2\%    & 3.58\%   &   2.32\% \\ \hline
\end{tabular}
\end{table}
\begin{table}[H]
\centering
\caption{The difference between the accuracy on the test samples (separable states, entangled states) from SSL or SLK and SL by using  $l$ labeled quantum states and $u$ unlabeled quantum states with $K=4,$ represented by
$\Delta_{SSL}$ $(\Delta_{SSL}^{-}, \Delta_{SSL}^{+})$ and $\Delta_{SLK}$ $(\Delta_{SLK}^-,\Delta_{SLK}^+)$.}
\label{tab2}
\begin{tabular}{|l|l|l|l|l|l|l|l|l|}
\hline
$l$  &u&$\Delta^{-}_{SSL}$&$\Delta^{-}_{SLK}$&$\Delta^{+}_{SSL}$&$\Delta^{+}_{SLK}$&$\Delta_{SSL}$&$\Delta_{SLK}$ \\ \hline
500  &10000 & 23.73\% & 25\%     & 2.83\%    & -4.83\% & 13.28\%  &  10.08\% \\ \hline
1000 &20000 & 6.7\%   & 5.37\%   & 9.93\%    & 7.87\%  & 8.32\%   &   6.62\% \\ \hline
2000 &40000 & 7.3\%   & 11.63\%  & 6.17\%    & -2.13\% & 6.75\%   &   4.75\% \\ \hline
4000 &80000 & 6.43\%  & 8.3\%    & 4.6\%     & -1.27\% & 5.52\%   &   3.52\% \\ \hline
\end{tabular}
\end{table}

When $K=4,$ the accuracy from SL, SSK and SSL can reach 76.23\% , 86.32\% and 89.52\%, respectively, by using $500$ labeled and $10000$ unlabeled quantum states, and 87.72\% , 91.23\% and 93.23\%, respectively, by using $4000$ labeled and $80000$ unlabeled quantum states. The accuracy via SSL by using 500 labeled quantum states and 10000 unlabeled states is higher than the one via SL using 4000 labeled states. In table \ref{tab1} and \ref{tab2}, we list the difference  between the accuracy on the test states from SSL or SLK and SL by using $l$ labeled and $u$ unlabeled data with $K=2$ and $4$. $\Delta^{-}_{SSL}$ ($\Delta^{+}_{SSL}$) is the difference  between accuracy on negative (positive) states from SSL  and SL. $\Delta^{-}_{SLK}$ ($\Delta^{+}_{SLK}$) is the difference between the accuracy  on negative (positive) states from SLK and SL.  $\Delta_{SSL}$ ($\Delta_{SLK}$) is the difference between accuracy on the test states by SSL (SLK) and SL. Here, negative and positive represent separable and entangled samples, respectively.

Accuracy and the Area Under Curve (AUC) of the Receiver Operating Characteristic (ROC) curve are used as performance measures. Let $TP$, $NP$, $P$ and $N$ be the numbers of positive and predicted as positive samples, negative but predicted as positive samples, total positives and total negatives, respectively. Let $TPR$ and $FPR$ be true positive (actual positive and predicted as positive) rate and the false positive (actual negative but predicted as positive) rate, respectively. We have
\begin{eqnarray}\label{TFPR}
\label{TP}
&TPR=\mathcal{P}(\hat P|P)=\frac{TP}{P},\\
&FPR=\mathcal{P}(\hat P|N)=\frac{NP}{N},\nonumber
\end{eqnarray}
where $\mathcal{P}(\hat P|X)$ is the posterior probability that the instances $X$ are positive. Now we specify a threshold value $\alpha.$  The output sample will be identified as a positive sample (an entangled state) if its class probability  by the classifier is greater than $\alpha$. Otherwise, it will be identified as a negative sample (a separable state). Then we can calculate $TPR$ and $FPR$ for a given $\alpha$. The ROC curve is a graph of $TPR$ as the $y$-axis and $FPR$ as the $x$-axis. The area under an ROC curve is a measure of the usefulness of a test in general (a larger area stands for a more useful test). The ROC curve and AUC of the test samples by SSL, SLK and SL are shown in FIG. \ref{2-roc-k2} and \ref{2-roc-k4} for $l=500(1000,2000, 4000$ and $u=10000(20000,40000,80000),$ $K=2$ or $K=4$.
The best ROC curves and AUC are attained via SSL for the same number of labeled data. The model trained by SSL has the best performance and the AUC reaches 0.983, which is greater than 0.948 via SL when $K=4,$ $l=4000$ and $u=80000$.
The accuracy and the AUC via SSL by using 500 labeled states and 10000 unlabeled states are almost equal to the ones via SLK by using 2000 labeled states when $K=2$, and are also higher than the ones via SL by using 4000 labeled states when $K=2$ and $K=4$.

\begin{figure}[H]
\centering
\includegraphics[width=8.5cm]{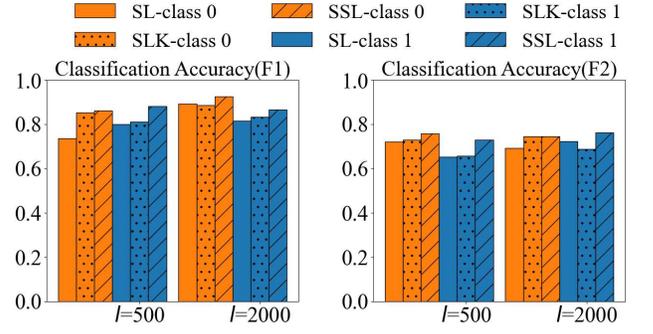}
\caption{Classification accuracy from SSL, SLK and SL  by using $l$ labeled and $u$ unlabeled quantum states when $K=4,$  $l=500,2000$ and $u=5000,20000$. Each set of three orange columns represents the classification accuracy on 500 or 2000 separable states (class-0) from SL, SLK and SSL, respectively; each set of three blue columns represents the classification accuracy on 500 or 2000 entangled states (class-1) from SL, SLK and SSL, respectively.}
\label{2-classa-F12}
\end{figure}

\begin{figure}[H]
\centering
\includegraphics[width=8.5cm]{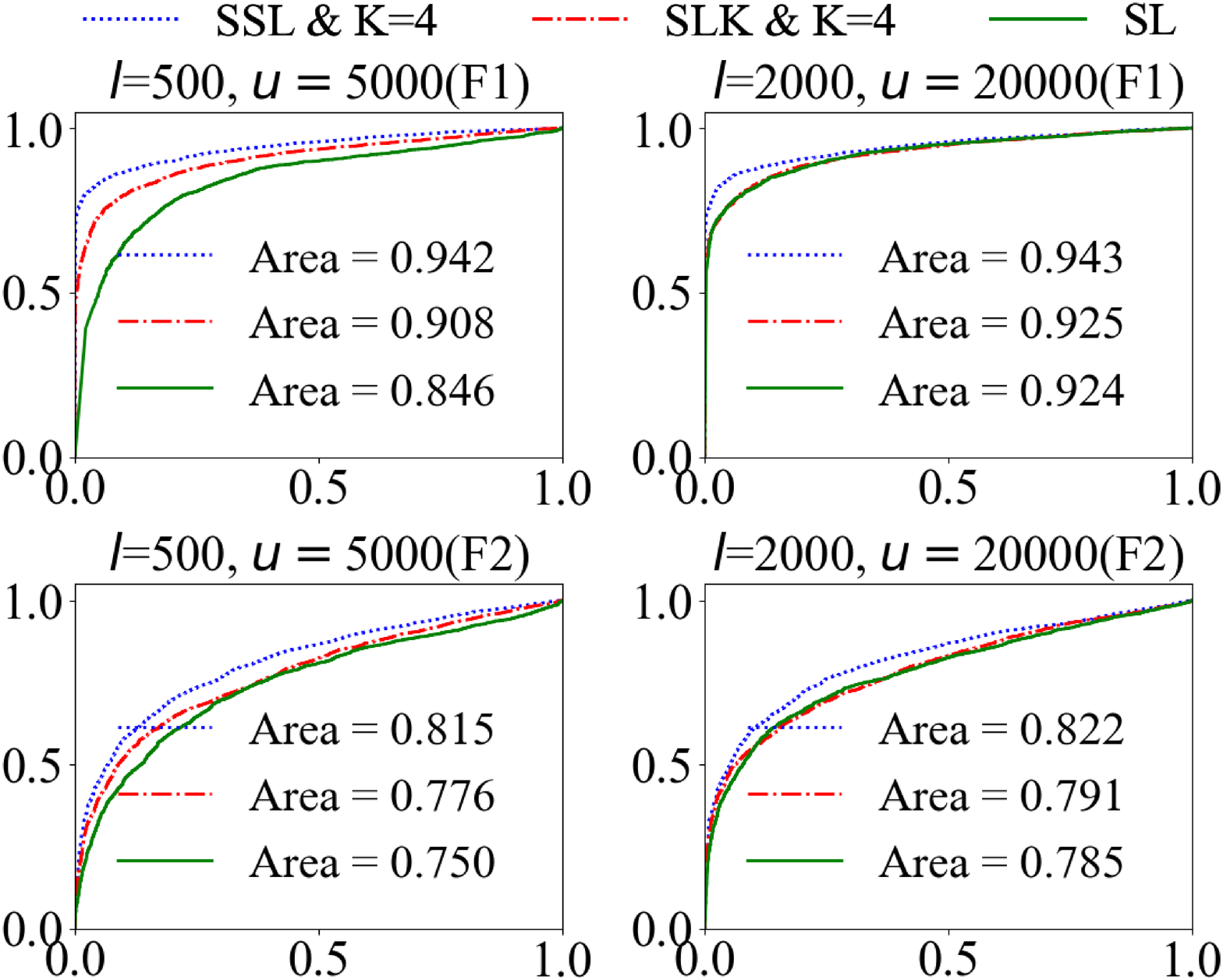}
\caption{The ROC curves of the test samples by SSL, SLK and SL are represented by the blue dotted line, red dash-dotted line and green solid line, respectively, for $l=500,2000$ and  $u=5000,20000$ and $K=4$, the horizontal axis represents the false positive rate and the longitudinal axis represents the positive rate.}
\label{2-ROC-F12}
\end{figure}

We show the classification accuracy and the ROC curves by using the partial information $F_1$ and $F_2$ as the feature vectors in FIG. \ref{2-classa-F12} and FIG. \ref{2-ROC-F12}. The classification accuracy can be improved a lot by semi-supervised machine learning. The accuracy using partial information $F_1$ is acceptable, but lower by using $F_2$ as the feature vectors. The accuracy via SSL by using $500$ labeled states and 10000 unlabeled states is higher than that from SLK by using 2000 labeled quantum states.

\begin{table}[H]
\centering
\caption{The accuracy on the test samples  from SSL, SLK and SL by using  $l$ labeled quantum states and $10l$ unlabeled quantum states with $K=4,$ represented by acc$_{SSL}$, acc$_{SLK}$ and acc$_{SL}$.}
\label{tab-F-1-2}
\begin{tabular}{|l|l|l|l|l|l|l|l|l|}
\hline
&$l$  &u&acc$_{SSL}$&acc$_{SLK}$&acc$_{SL}$ \\ \hline
$F_1$&500  &5000 & 0.872 & 0.831    & 0.768    \\ \hline
$F_1$ &2000  &20000 & 0.895 & 0.859    & 0.854   \\ \hline
$F_2$&500  &5000 & 0.744 & 0.694    & 0.686    \\ \hline
$F_2$ &2000  &20000 & 0.754 & 0.717    & 0.707    \\ \hline
\end{tabular}
\end{table}

To test the performance of the semi-supervised model on the special case $\rho_s=\frac{1-p}{2}\rm{I}_2\otimes \rho_B+p|\psi\rangle\langle\psi|$ with $|\psi\rangle=\cos\theta|00\rangle+\sin\theta|11\rangle,$ $\rho_B=\rm{tr}_A[|\psi\rangle\langle\psi|]$
and $\theta=\frac{\pi}{8},$ we use $l$ labeled states (general 2-qubit quantum states) and $2l$ unlabeled states $\rho_s$ as the training set, 2000 states $\rho_s$ ($1000$ separable states and $1000$ entangled states) as the test set. We list the accuracy using the whole information $F$ and partial information $F_1$ and $F_2$ as the feature vectors in TABLE \ref{rhos-F}.  The accuracy can reach about 0.84 by supervised machine learning, and it can be improved to be about 0.97 by SSL when we only use 30 labeled states and 60 unlabeled states. Interestingly, the accuracy by using the partial information $F_2$ as the feature vectors is higher than that by using the whole information as the feature vectors. The accuracy via semi-supervised model by using only 30 labeled states and 60 unlabeled states is higher than that from
SLK by using 100 labeled states when $F$ and $F_1$ are the feature vectors.

\begin{table}[H]
\centering
\caption{The accuracy on the test samples $\rho_s$ (1000 separable states, 1000 entangled states) from SSL or SLK and SL by using  $l$ labeled quantum states and $2l$ unlabeled quantum states $\rho_s$ with $K=4,$ represented by acc$_{SSL}$, acc$_{SLK}$ and acc$_{SL}$.}
\label{rhos-F}
\begin{tabular}{|l|l|l|l|l|l|l|l|l|}
\hline
&$l$  &u&acc$_{SSL}$&acc$_{SLK}$&acc$_{SL}$ \\ \hline
$F$&30  &60 & 0.972 & 0.717    & 0.841     \\ \hline
$F$&100  &200 & 0.988 & 0.900    & 0.876     \\ \hline
$F_1$&30  &60 & 0.940 & 0.871    & 0.795     \\ \hline
$F_1$&100  &200 & 0.987 & 0.900    & 0.849     \\ \hline
$F_2$&30  &60 & 0.976 & 0.928    & 0.846    \\ \hline
$F_2$&100  &200 & 0.992 & 0.990    & 0.965    \\ \hline
\end{tabular}
\end{table}

\subsection{Classifying three-qubit GHZ states mixed with white noise}
\begin{figure}[H]
\centering
\includegraphics[width=8.5cm]{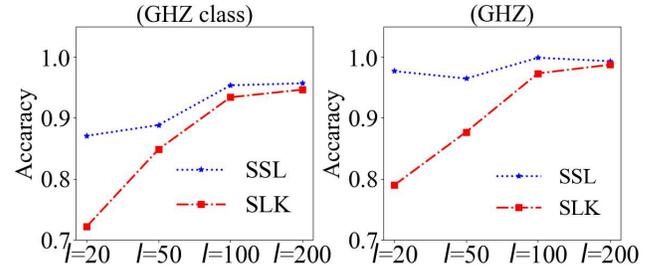}
\caption{Classification accuracy on the noisy three-qubit GHZ class states (left) and the noisy three-qubit GHZ states (right) as the test samples via SSL and SLK is represented by the blue dotted lines with $\star$ and red dash-dotted lines with $\Box$ for $l=20,50,100$ and $200$ and $u=50,150,1000,2000$.}
\label{3-accuracy}
\end{figure}

\begin{figure}[H]
\centering
\includegraphics[width=8.5cm]{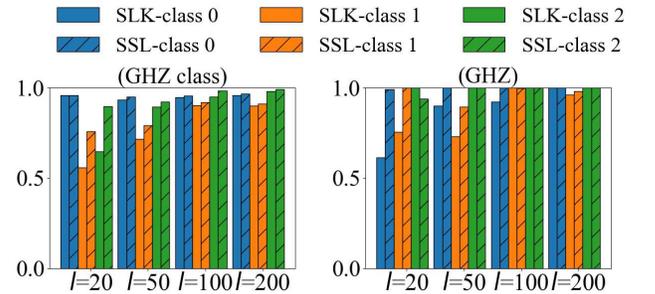}
\caption{Classification accuracy on the noisy three-qubit GHZ class states (left) and the noisy three-qubit GHZ states (right) as the test samples via SSL and SLK when $K_1=2$, $K_2=8$, $l=20,50,100,200$ and $u=50,150,1000,2000$. The first (second) blue column represents the classification accuracy on $2000$ separable states (class-0) by SLK (SSL); the third (fourth) orange column represents the classification accuracy on $2000$ biseparable states (class-1) by SLK (SSL); the fifth (sixth) green column represents the classification accuracy on  $2000$ entangled states (class-2) by SLK (SSL).}
\label{3-classa}
\end{figure}

\begin{figure}[H]
\centering
\includegraphics[width=8.5cm]{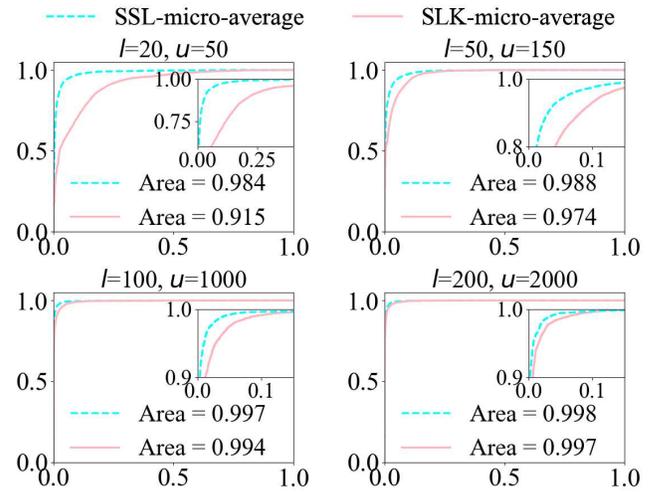}
\caption{ The micro-averaged ROC curves of the test samples by SSL and SLK are represented by the cyan dashed line and pink solid line, respectively, when $K_1=2$, $K_2=8,$ $l=20,50,100, 200$ and $u=50,150,1000,2000$, the horizontal axis represents the false positive rate and the longitudinal axis represents the positive rate.}
\label{3-micro-average}
\end{figure}

\begin{figure}[H]
\centering
\includegraphics[width=8.5cm]{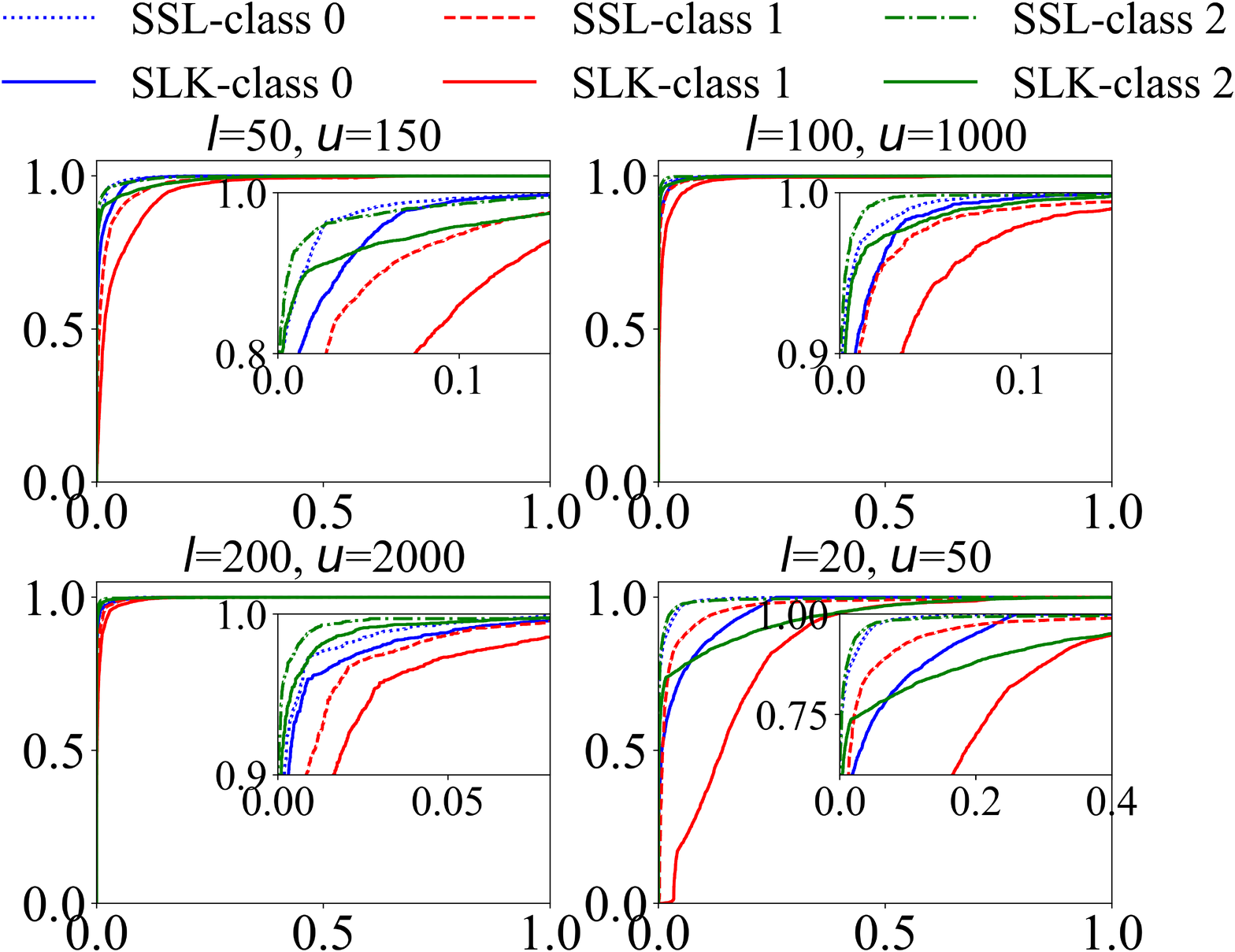}
\caption{The ROC curves of the test samples by SSL and SLK are represented by the dashed lines and the solid lines, respectively, when $K_1=2$, $K_2=8,$ $l=20,50,100,200$ and $u=50,150,1000,2000$, the horizontal axis represents the false positive rate and the longitudinal axis represents the positive rate. The area of class-0, class-1 and class-2 by SSL (SLK) is: (1) when $l=20,u=50,$ 0.993 (0.958), 0.969 (0.832) and 0.992 (0.937),
(2) when $l=50,u=150$, 0.995 (0.990), 0.977 (0.954) and 0.995 (0.988),  (3) when $l=100,u=1000$, 0.998 (0.997), 0.994 (0.986) and 0.999 (0.997), (4) when $l=200,u=2000$, 0.999 (0.998), 0.996 (0.993) and 0.999 (0.999), respectively.}
\label{3-class-roc}
\end{figure}

Different from the classification of two-qubit quantum states, $K_1(K_2)$ augmentations are performed on the labeled (unlabeled) states and $K_1\neq K_2$ for classifying three-qubit noisy GHZ states. The noisy $3$-qubit GHZ class states are attained when local unitary operations are performed on three-qubit GHZ state mixed with white noise. We have two sets of test samples. One set consists of $2000$ separable, $2000$ biseparable and $2000$ entangled noisy three-qubit GHZ states. Another set consists of $2000$ separable, $2000$ biseparable and $2000$ entangled noisy three-qubit GHZ class states. The accuracy on the two test sets are computed by SSL and SLK by using $l$ labeled and $u$ unlabeled quantum states with $K_1=2$, $K_2=8,$ $l=20,50,100,$ $200$ and $u=50,150,1000,2000$. The numerical results are shown in FIGs. \ref{3-accuracy}, \ref{3-classa}, \ref{3-micro-average} and \ref{3-class-roc}.

As shown in the FIG. \ref{3-accuracy} and FIG. \ref{3-classa}, the accuracy of the noisy three-qubit GHZ class states by the SLK and SSL models is 72.18\% and 87.08\%, and the one of the noisy three-qubit GHZ states is 78.98\% and 97.7\%, respectively, for $l=20$ and $u=50$. The accuracy of the noisy three-qubit GHZ class states by the SLK and SSL models is 94.66\% and 95.7\%, and the one of the noisy three-qubit GHZ states is 98.78\% and 99.33\%, respectively, for $l=200$ and $u=2000$.
In TABLE \ref{tab3} and \ref{tab4}, we list the difference between the accuracy on positive (negative) states from SKL and SSL by using $l$ labeled and $u$ unlabeled data. $\Delta^{0}_{SSL}$, $\Delta^{1}_{SSL}$ and $\Delta^{2}_{SSL}$ represent the difference between the accuracy on class-0 (separable states), class-1 ( biseparable states) and class-2 (entangled states) by SSL and SLK, respectively. $\Delta_{SSL}$ is the difference between accuracy on all test samples by SSL and SLK. In table \ref{tab4}, the difference $\Delta^{2}_{SSL}$ is 0 since the classification accuracy of class-2 by SSL and SLK are all 100\%. Compared with the results from SLK, the accuracy of SSL is higher in most cases.
\begin{table}[H]
\centering
\caption{The difference between the accuracy on the noisy three-qubit GHZ class samples from SSL and SLK by using $l$ labeled quantum states and $u$ unlabeled quantum states with $K_1=2$ and $K_2=8$.}
\label{tab3}
\begin{tabular}{|l|l|l|l|l|l|}
\hline
$l$ &u&$\Delta_{SSL}$ &$\Delta^{0}_{SSL}$&$\Delta^{1}_{SSL}$&$\Delta^{2}_{SSL}$\\ \hline
20  &50  & 14.9\%   & -0.2\%    & 19.85\%  & 25.05\%  \\ \hline
50  &150  & 3.98\%   & 1.33\%    & 2.18\%  & 1.08\%   \\ \hline
100 &1000 & 1.96\%   & 0.88\%    & 1.7\%   & 3.3\%  \\ \hline
200 &2000 & 1.04\%   & 0.88\%    & 1.15\%  & 1.1\%  \\ \hline
\end{tabular}
\end{table}
\begin{table}[H]
\centering
\caption{The difference between the accuracy on the noisy three-qubit GHZ samples from SSL and SLK by using $l$ labeled quantum states and $u$ unlabeled quantum states with $K_1=2$ and $K_2=8$.}
\label{tab4}
\begin{tabular}{|l|l|l|l|l|l|}
\hline
$l$ &u&$\Delta_{SSL}$ &$\Delta^{0}_{SSL}$&$\Delta^{1}_{SSL}$&$\Delta^{2}_{SSL}$\\ \hline
20  &50  & 18.72\%   & 37.6\%    & 24.55\%    & -6\%  \\ \hline
50  &150  & 8.77\%    & 9.95\%   & 16.35\%      & 0\%   \\ \hline
100 &1000 & 2.57\%   & 7.9\%    & -0.2\%   & 0\%  \\ \hline
200 &2000 & 5.83\%   & 0\%      & 1.75\%   & 0\%  \\ \hline
\end{tabular}
\end{table}
The micro-averaged ROC curves are used as the performance measures for the classification problem for the noisy three-qubit GHZ states. The idea is to split the three-class classification problem into three two-class classification problems, and calculate the arithmetic mean of $TP$, $NP$, $P$ and $N$ attained from the three classification problems. Then using Eq. \eqref{TP}, the micro-averaged ROC curves can be attained as shown in FIG. \ref{3-micro-average}. And the ROC curves of the class-0 (separable states), class-1 ( biseparable states) and class-2 (entangled states) by SLK and SSL are shown in FIG. \ref{3-class-roc}. Obviously, the best classifiers obtained by SSL.

\subsection{Classifying n-qubit GHZ states mixed with white noise}

\begin{figure}[H]
\centering
\includegraphics[width=8.5cm]{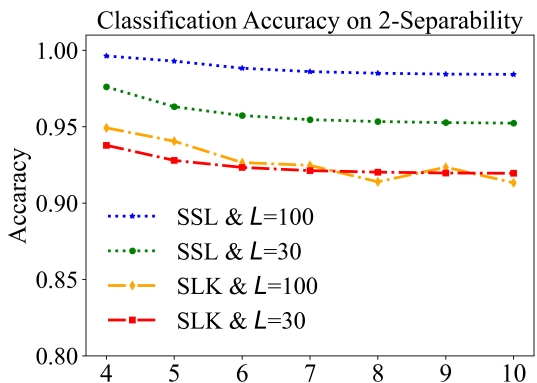}
\caption{Classification accuracy on the 2-separable states and genuine multipartite entangled states of the noisy n-qubit GHZ states via SSL and SLK when $K_1=K_2=1$, $l=30,100$ and $u=2l$. The accuracy via SSL is represented by green dotted line with $\bullet$ when $l=30$
and blue dotted line with $\star$ when $l=100$, the accuracy via SLK is represented by red dash dotted line with the square when $l=30$ and orange dash dotted line with $\diamond$ when $l=100$.}
\label{n-qubit-2s}
\end{figure}

\begin{figure}[H]
\centering
\includegraphics[width=8.5cm]{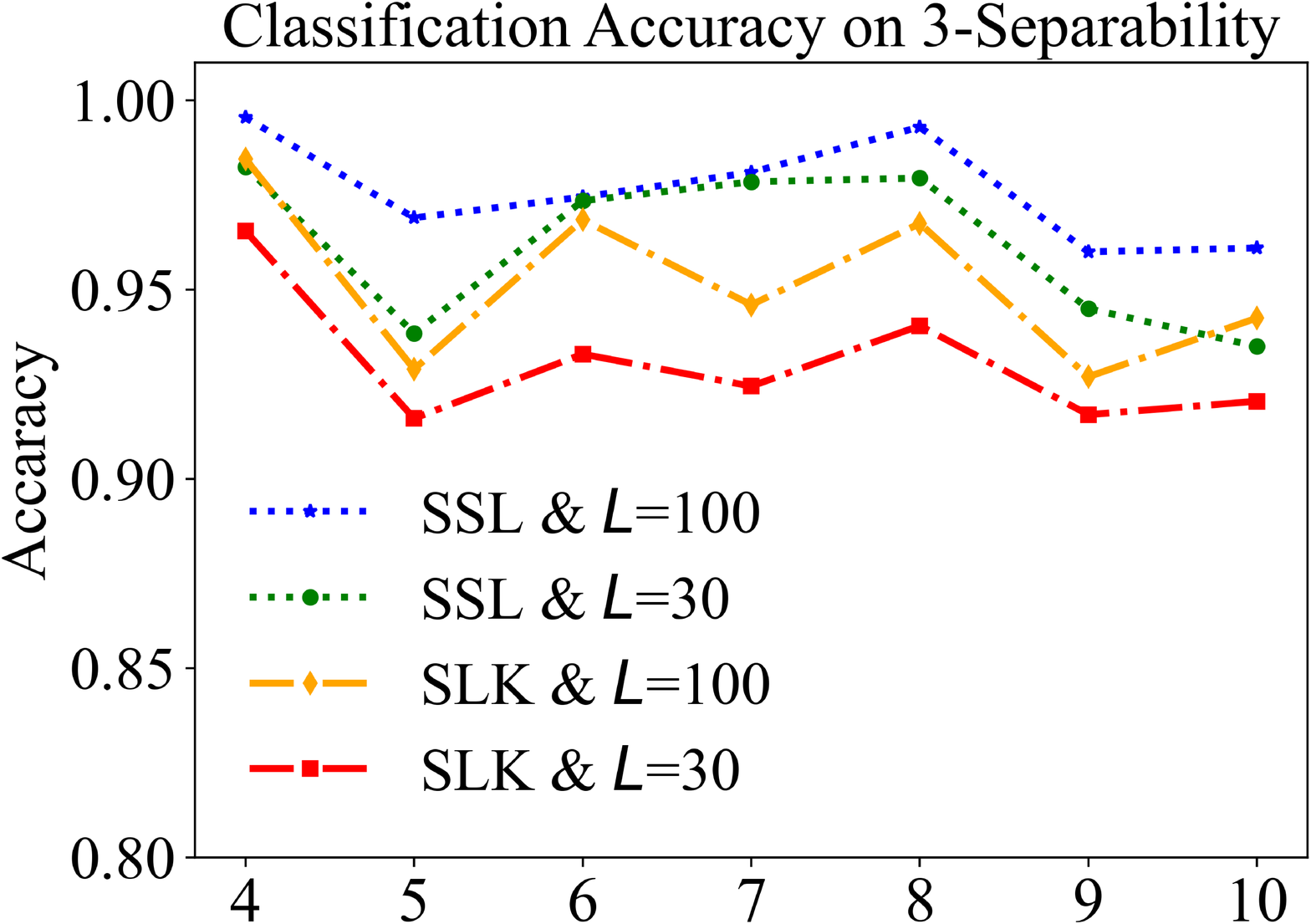}
\caption{Classification accuracy on the 3-separable states and 3-nonseparable states of the noisy n-qubit GHZ states via SSL and SLK when $K_1=K_2=1$, $l=30,100$ and $u=2l$. The accuracy via SSL is represented by green dotted line with $\bullet$ when $l=30$
and blue dotted line with $\star$ when $l=100$, the accuracy via SLK is represented by red dash dotted line with the square when $l=30$ and orange dash dotted line with $\diamond$ when $l=100$.}
\label{n-qubit-3s}
\end{figure}

\begin{figure}[H]
\centering
\includegraphics[width=8.5cm]{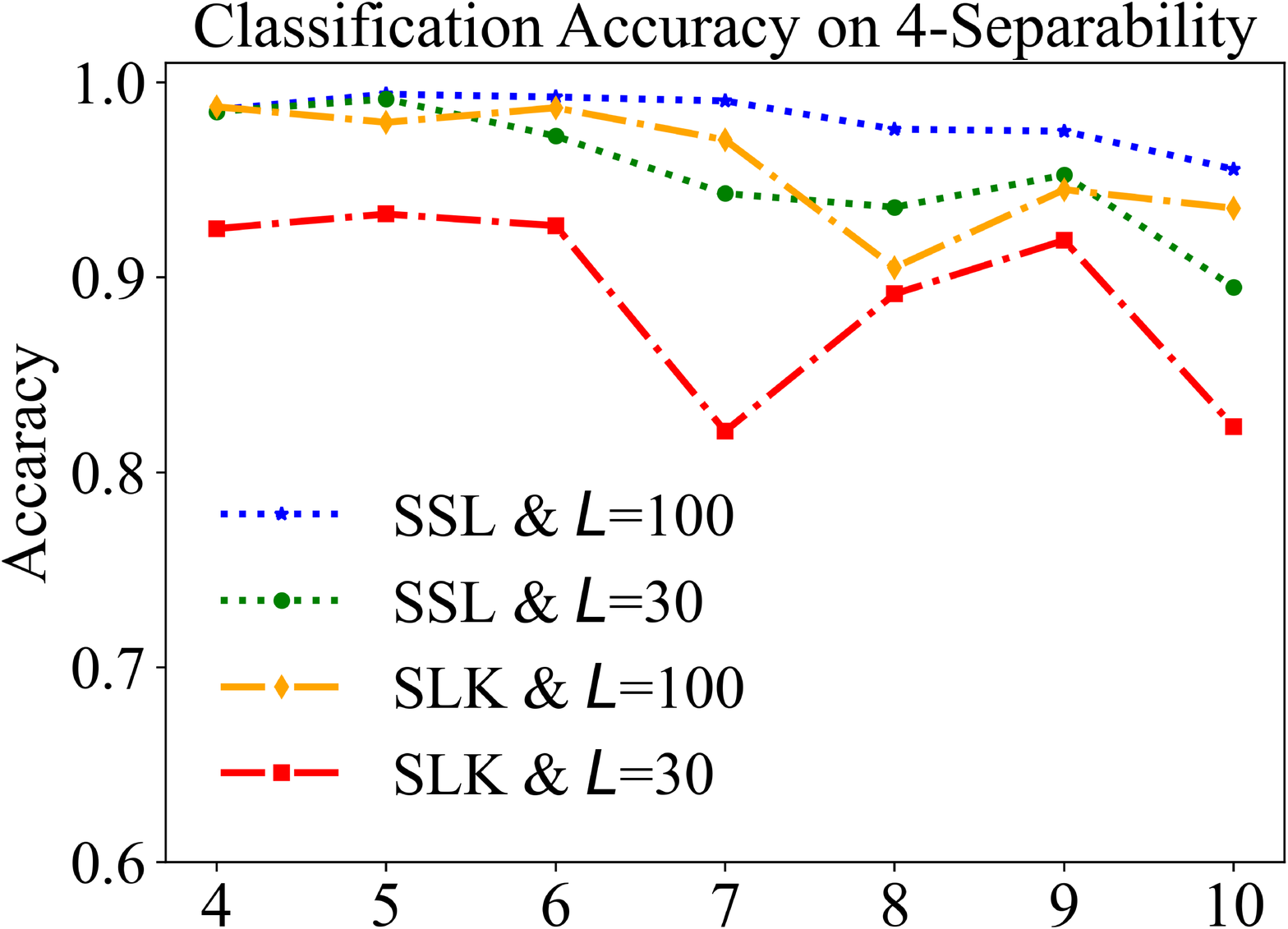}
\caption{Classification accuracy on the 4-separable states and 4-nonseparable states of the noisy n-qubit GHZ states via SSL and SLK when $K_1=K_2=1$, $l=30,100$ and $u=2l$. The accuracy via SSL is represented by green dotted line with $\bullet$ when $l=30$
and blue dotted line with $\star$ when $l=100$, the accuracy via SLK is represented by red dash dotted line with the square when $l=30$ and orange dash dotted line with $\diamond$ when $l=100$.}
\label{n-qubit-4s}
\end{figure}

To test the performance of semi-supervised model on $n$-qubit states, we train a model of $k$-separability $(k=2,3,4)$ for the $n$-qubit noisy GHZ states $(n=4,\cdots,10)$.
We generate $l$ labeled states $\rho_{ng}$ and $2l$ unlabeled states, $K_1=K_2=1$ data augmentation technique is applied by using the Pauli matrices. The feature vectors are $\{\langle M_x\rangle,\langle M_z\rangle\}$ with $M_x=\sigma_x^{\otimes n}$
and $M_z=|0\rangle\langle 0|^{\otimes n}+|1\rangle\langle 1|^{\otimes n}$. The accuracy and the difference of the accuracy are shown in FIGs.\ref{n-qubit-2s}, \ref{n-qubit-3s} and \ref{n-qubit-4s} when $l=30$ and $100$, and in TABLEs. \ref{ghz-2s}, \ref{ghz-3s} and \ref{ghz-4s}, where $\Delta=acc_{SSL}-acc_{SLK}$. The accuracy via SSL using 30 labeled states and 60 ublabeled states is higher than the one via SLK using 100 labeled states in most cases.

\begin{table}[H]
\centering
\caption{The accuracy and the difference between the accuracy on the 2-separable states and genuine multipartite entangled states of the noisy n-qubit GHZ samples from SSL and SLK by using $l$ labeled quantum states and $u=2l$ unlabeled quantum states with $K_1=K_2=1$.}
\label{ghz-2s}
\begin{tabular}{|l|l|l|l|l|l|l|l|l|}
\hline
&$l$ &$n=4$ &$n=5$ &$n=6$&$n=7$ &$n=8$ &$n=9$ &$n=10$ \\ \hline
SLK&100  & 0.949 & 0.941 & 0.927 & 0.925&  0.914 & 0.924 & 0.913   \\ \hline
SSL&100  & 0.996 & 0.993 & 0.988 & 0.986&  0.985 & 0.985 & 0.984  \\ \hline
$\Delta$&100&  0.047 &  0.053 & 0.062 & 0.061&  0.071 &0.061&  0.071 \\ \hline
SLK&30 & 0.938  &  0.928 & 0.923 & 0.921&  0.920 &0.920& 0.920 \\ \hline
SSL&30 & 0.976 & 0.963 & 0.957 & 0.955 &   0.953 &0.953 & 0.952\\ \hline
$\Delta$&30&  0.038&0.035 &  0.034 & 0.034 &  0.033 &  0.033& 0.032 \\ \hline
\end{tabular}
\end{table}

\begin{table}[H]
\centering
\caption{The accuracy and the difference between the accuracy on the 3-separable states and 3-nonseparable states of the noisy n-qubit GHZ samples from SSL and SLK by using $l$ labeled quantum states and $u=2l$ unlabeled quantum states with $K_1=K_2=1$.}
\label{ghz-3s}
\begin{tabular}{|l|l|l|l|l|l|l|l|l|}
\hline
&$l$ &$n=4$ &$n=5$ &$n=6$&$n=7$ &$n=8$ &$n=9$ &$n=10$ \\ \hline
SLK&100  & 0.985 & 0.929 & 0.969 & 0.946&  0.968 & 0.927 & 0.943   \\ \hline
SSL&100  & 0.996 & 0.969 & 0.975 & 0.981&  0.993 & 0.96 & 0.961  \\ \hline
$\Delta$&100&  0.011 &  0.04 & 0.006 & 0.035&  0.026 &0.033&  0.018 \\ \hline
SLK&30 & 0.966  &  0.916 & 0.933 & 0.925&  0.941 &0.917& 0.921 \\ \hline
SSL&30 & 0.983 & 0.939 & 0.974 & 0.979 &   0.980 &0.945 & 0.935\\ \hline
$\Delta$&30&  0.017&0.023 &  0.041 & 0.054 &  0.039 &  0.028& 0.014 \\ \hline
\end{tabular}
\end{table}

\begin{table}[H]
\centering
\caption{The accuracy and the difference between the accuracy on the 4-separable states and 4-nonseparable states of the noisy n-qubit GHZ samples from SSL and SLK by using $l$ labeled quantum states and $u=2l$ unlabeled quantum states with $K_1=K_2=1$.}
\label{ghz-4s}
\begin{tabular}{|l|l|l|l|l|l|l|l|l|}
\hline
&$l$ &$n=4$ &$n=5$ &$n=6$&$n=7$ &$n=8$ &$n=9$ &$n=10$ \\ \hline
SLK&100  & 0.988 & 0.980 & 0.987 & 0.971&  0.905 & 0.945 & 0.936   \\ \hline
SSL&100  & 0.986 & 0.994 & 0.993 & 0.991&  0.976 & 0.975 & 0.956  \\ \hline
$\Delta$&100& -0.002 &  0.015 & 0.006 & 0.02&  0.071 &0.03&  0.02 \\ \hline
SLK&30 & 0.925  &  0.933 & 0.927 & 0.821&  0.892 &0.919& 0.824 \\ \hline
SSL&30 & 0.985 & 0.992 & 0.973 & 0.943 &   0.936 &0.953 & 0.895\\ \hline
$\Delta$&30&  0.06&0.059 &  0.046 & 0.122 &  0.045 &  0.034& 0.072 \\ \hline
\end{tabular}
\end{table}

The necessary and sufficient conditions of the $k$-separability for the state $\rho_{ng}$ are known when $k=2$ and $k\geq \frac{n+1}{2}.$
The analytical conditions of $3$-separability of $\rho_{ng}$ are still lacking when $n=6$ and $7.$
We will employ the SL and SSL to give the predictions of $3$-separability for $\rho_{ng}$ for $n=4,5,6$ and $7.$
To show the validity of the SL and SSL, the predicted white noise tolerances for $3$-separability
of $\rho_{ng}$  are compared with the necessary and sufficient analytical bounds when $n=4$ and $5$, and  the numerical values from the linear programming method  when $n=6$ and $7$\cite{p2}.

We employ the SSL method to detect the $3$-separability of $\rho_{ng}$ by using 200 labeled and 1000 unlabeled noisy GHZ states.  To obtain the predictions  of $3$-separability and avoid perturbations of other bounds by SLK and SSL, we will train the classifiers as a binary classification problem($3$-separability and $3$-nonseparability) similar to  the entanglement detection problem of two-qubit states.  As we do not know
exactly the precise bounds of $3$-separability for $n=6$ and $n=7,$ we will generate the labeled and test states from some fuzzy bounds based on $4$-qubit and $5$-qubit noisy GHZ states, i.e.
the volume of $3$-separable($3$-nonseparable) states accounts for $\frac{1}{4}$ (about $0.641$) of the total volume of $2$-separable but $4$-nonseparable states for $4$-qubit($5$-qubit) noisy GHZ states.

The class of $3$-separability is generated randomly from $0\le p\le b_4+a\frac{(b_2-b_4)}{4},$ the one of $3$-nonseparability is generated from $b_2-a\frac{2(b_2-b_4)}{3}\le p\le 1,$ where $a$ is a parameter and $0<a<1$. To balance the unlabeled states, the convex combination of the unlabeled states are performed. Then
the augmentation strategies by performing local unitary transformations are employed on the labeled and unlabeled states five times, and the test states once.
We only use the augmented states for training, validation and testing.

In the numerical experiment, 2000 test samples are generated  from $p=0$ to $p=1$ with the step length $h=0.0005$. Then the test samples are predicted via the model from $p=0$ to 1. Ten quantum  states exist in each interval $[10n_1h,10(n_1+1)h]$ with $n_1=0,1,\cdots, 199.$  We set $b_3$ as the midpoint of the first interval such that the number of  3-nonseparable states is five or greater than five from $n_1=0$ to 199. The bounds of $3$-separability learned by SLK and SSL  are shown  in  FIG. \ref{b_7_8}($a=\frac{7}{8}$), FIG. \ref{b_3_4}($a=\frac{3}{4}$) and FIG. \ref{b_1_2}($a=\frac{1}{2}$), which are attained by  using eight different sets of 200 labeled(200 labeled and 1000 unlabeled) states for $n=4,\cdots,7$. Obviously, the bounds predicted by the SSL approach coincide with the known bounds quite well.
\begin{figure}[H]
\centering
\includegraphics[width=8.5cm]{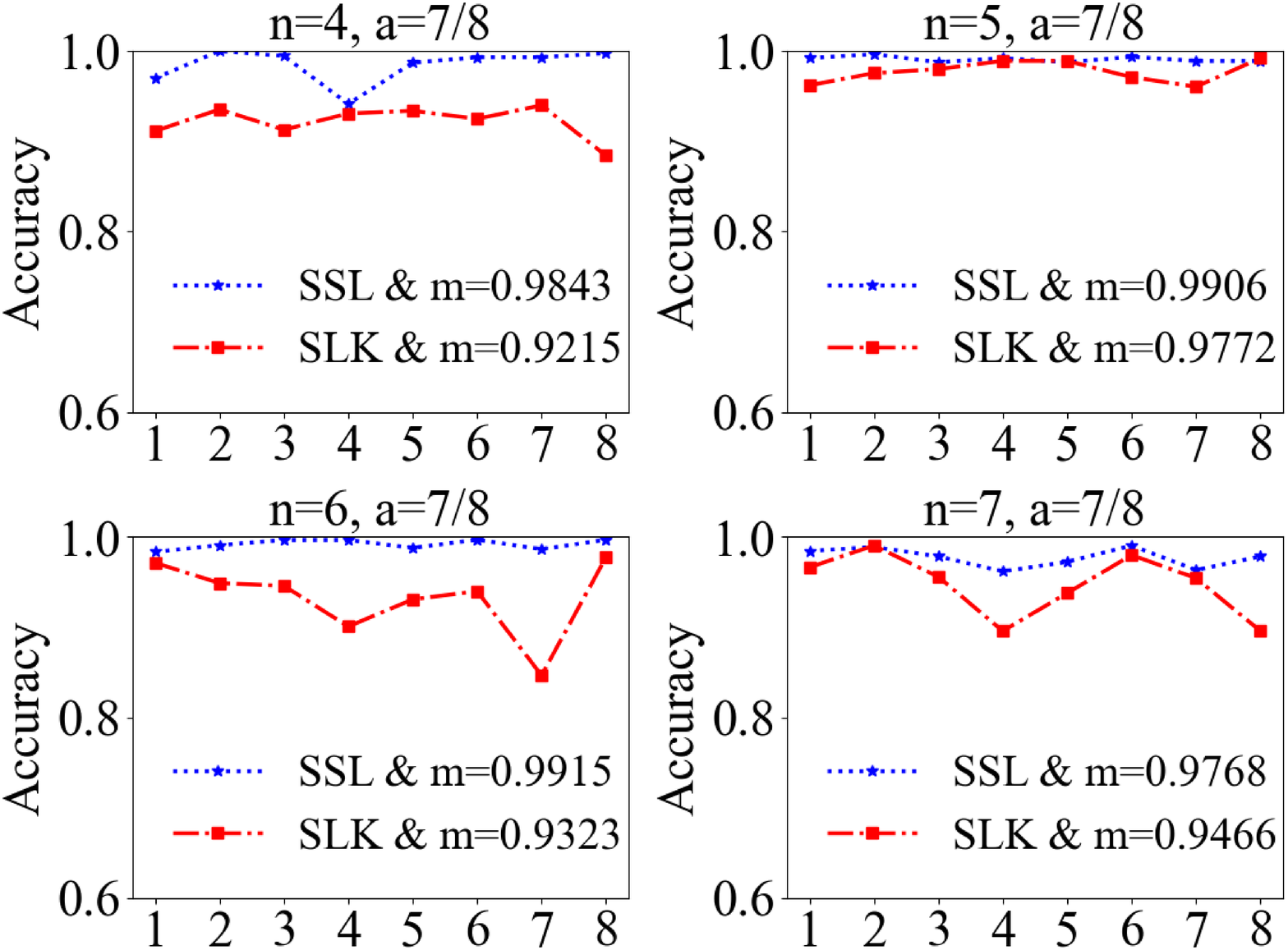}
\caption{Classification accuracy on the test samples via SSL and SLK are represented by the blue dotted lines with $\star$ and red dash-dotted lines with $\Box$ for
$l=200$ and $u=1000$, the horizontal axis represents different training sets and the longitudinal axis represents the classification accuracy.}
\label{n-qubit-accuracy}
\end{figure}
\begin{figure}[H]
\centering
\includegraphics[width=8.5cm]{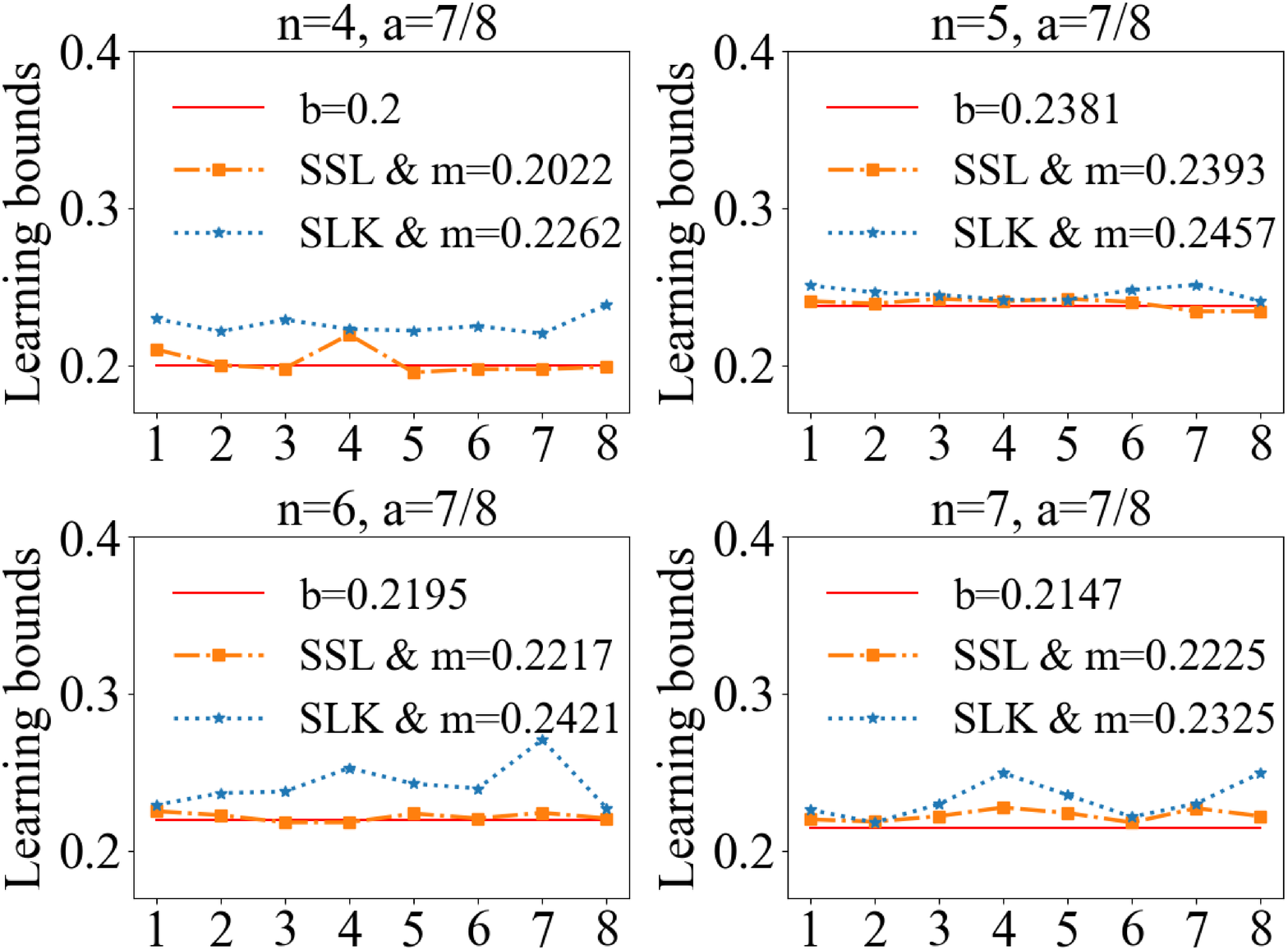}
\caption{The bounds of $3$-separability learned by SLK(the blue dotted lines with $\star$) and SSL(the orange dash-dotted lines with $\Box$)  using eight different sets of 200 labeled and 1000 unlabeled n-qubit noisy GHZ states for $a=\frac{7}{8}$ and $n=4,5,6$ and 7, the horizontal axis represents different training sets and the longitudinal axis represents $b$.}
\label{b_7_8}
\end{figure}
\begin{figure}[H]
\centering
\includegraphics[width=8cm]{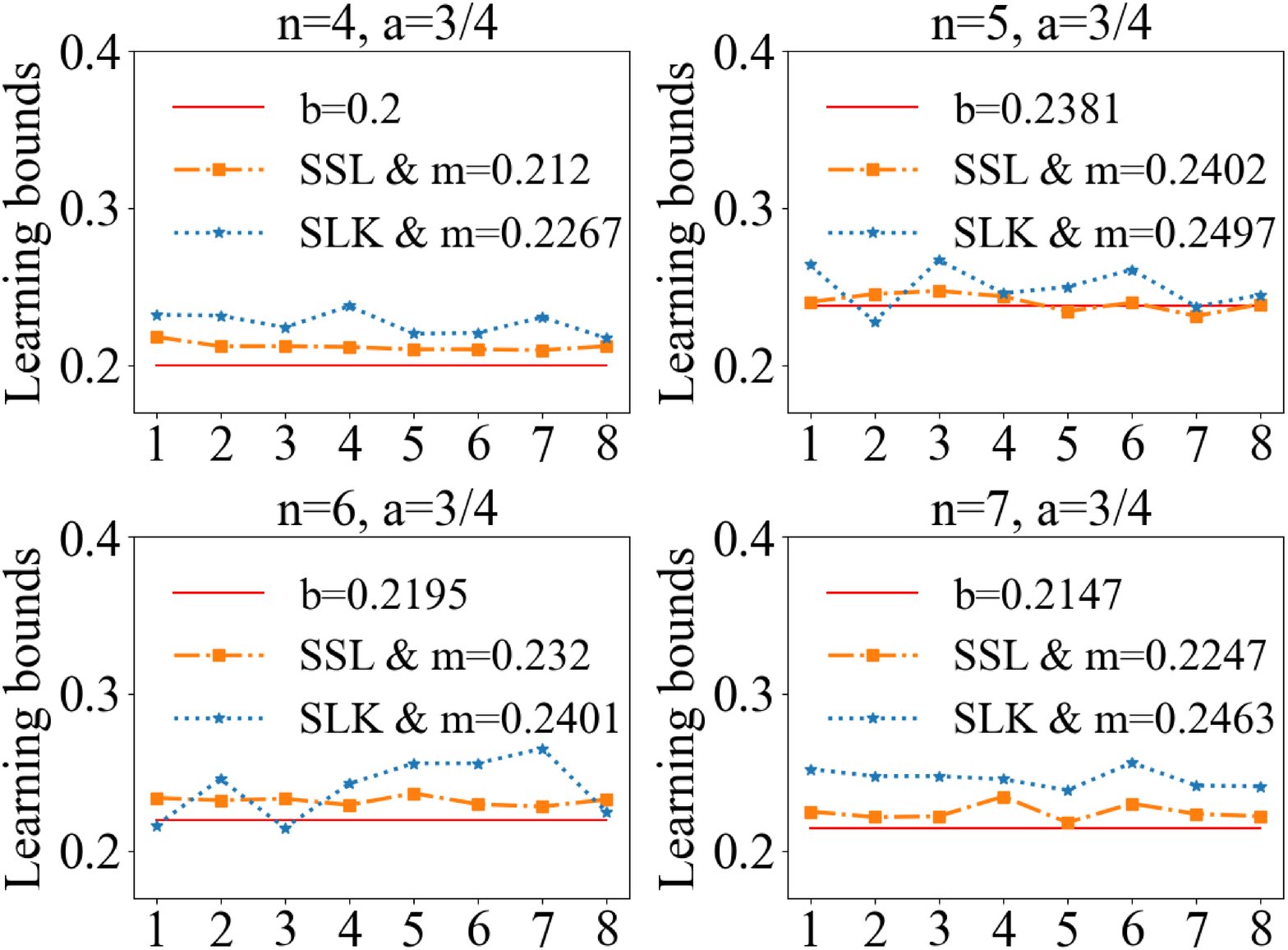}
\caption{The bounds of $3$-separability learned by SKL(the blue dotted lines with $\star$) and SSL(the orange dash-dotted lines with $\Box$)  using eight different sets of n-qubit noisy GHZ states for $a=\frac{3}{4}$ and $n=4,5,6$ and 7, the horizontal axis represents different training sets and the longitudinal axis represents $b$.}
\label{b_3_4}
\end{figure}
\begin{figure}[H]
\centering
\includegraphics[width=8cm]{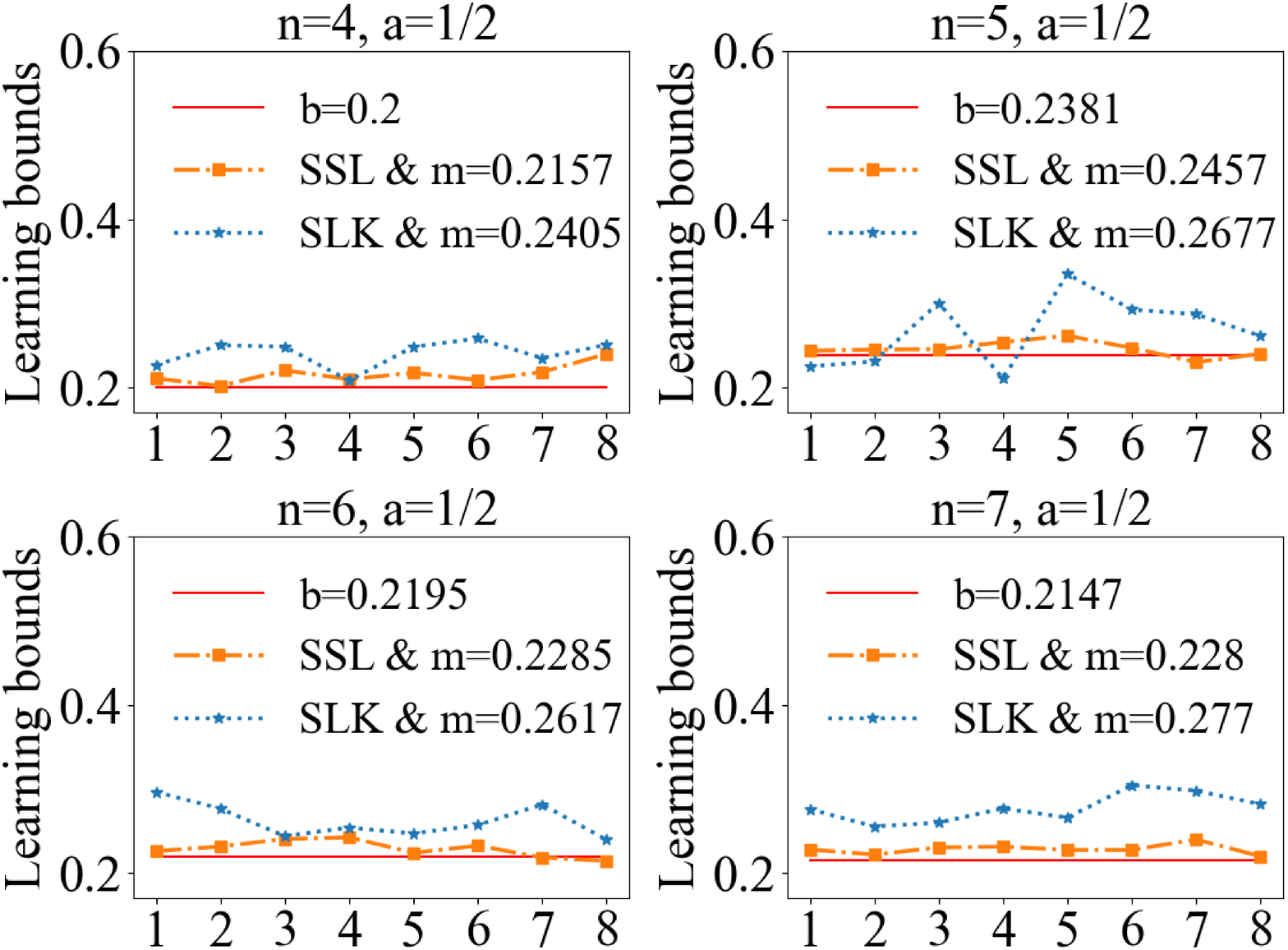}
\caption{The bounds of of $3$-separability by SLK(the blue dotted lines with $\star$) and SSL(the orange dash-dotted lines with $\Box$)  using eight different sets of n-qubit noisy GHZ states for $a=\frac{1}{2}$ and $n=4,5,6$ and 7, the horizontal axis represents different training sets and the longitudinal axis represents $b$.}
\label{b_1_2}
\end{figure}
The average accuracy becomes higher as $a$ becomes larger as the bounds are close to the optimal ones when $a$ becomes larger. We list the relative error of the average bounds by SSL and SLK compared to the known bounds in TABLE \ref{tab5} for $a=\frac{7}{8}$. The error of the predictions by SSL is smaller.
\begin{table}[H]
\centering
\caption{The average bounds  of $3$-separability of n-qubit noisy GHZ states learned by SLK and SSL ($a=\frac{7}{8}$) are represented by $b_{slk}$ and $b_{ssl}$ respectively, and the relative error of $b_{slk}$ and $b_{ssl}$ compared to $b_3$ is represented by $\rm{RE}_{slk}$ and $\rm{RE}_{ssl}$  respectively.}
\label{tab5}
\begin{tabular}{|l|l|l|l|l|l|}
\hline
n&$b_3$&$b_{slk}$&$b_{ssl}$&$\rm{RE}_{slk}$ &$\rm{RE}_{ssl}$\\ \hline
4&0.2    & 0.2262 & 0.2022   &13.09\% & 2.61\%     \\ \hline
5&0.2381 & 0.2457 & 0.2393   &3.20\% &1.31\%      \\ \hline
6&0.2195 & 0.2421 &  0.2217  &10.29\% &1.30\%      \\ \hline
7&0.2147 &0.2326  & 0.2225   &8.29\% &3.61\%      \\ \hline
\end{tabular}
\end{table}
Although the hyper-parameters, constant parameters and networks are usually given empirically, the results by semi-supervised machine learning are still better compared with supervised learning method for the entanglement detection problems of two-qubit and $n$-qubit noisy GHZ quantum states. In \cite{svm2} the BCHM classifier, which is the combination of supervised learning and the convex hull membership (CHM), has the advantage in both accuracy and speed. Compared with the supervised machine learning with $5\times 10^4$ training data in \cite{svm2}, the accuracy is higher by using only $4000$ labeled quantum states via our semi-supervised method. Our approach can also be combined with the CHM in \cite{svm2} to improve the accuracy further. The three-qubit noisy GHZ state, fully separable states, bi-separable states and genuine tripartite entangled states can be classified simultaneously, and the three-qubit GHZ class state can be also classified by using three-qubit noisy GHZ states as the training set via augmentation techniques.  Besides we trained a model of $k$-separability for the $n$-qubit noisy GHZ states from $n=4$ to $10$ $(k=2,3,4)$. The accuracy via SSL using 30 labeled and 60 unlabeled quantum states is higher than the one via SLK using 100 labeled quantum states in most cases. The predictions of $3$-separability of $n$-qubit noisy GHZ states when $n=4,5,6,7$ are learned by SSL using only the augmented states and the labeled states generated through the fuzzy bounds, which coincide with the known bounds well. The bounds of $k$-separability when $k\neq 3$ can also be predicted by the SLK and SSL methods. In the experiment, the partial information $F_1$ or $F_2$ of 2-qubit states and  $\{\langle M_x\rangle, \langle M_z\rangle\}$ of n-qubit noisy GHZ states can be attained instead of state tomography. Then the 2-qubit quantum states or the n-qubit noisy GHZ states can be classified by the semi-supervised model and the partial information, which may reduce the time complexity.

\section{Conclusion}
We have presented an efficient classifiers by employing semi-supervised machine learning method based on FixMatch and Pseudo-Label together with data augmentation for the entanglement detection of two-qubit and $n$-qubit quantum states. For three-qubit systems, the fully separable states, bi-separable states and genuine tripartite states can be classified simultaneously. Compared with the supervised machine learning, our detection accuracy is higher even the hyper-parameters are just given empirically.
For $n$-qubit($n=4,\cdots,7$) noisy GHZ states, although the labeled states are just generated by the fuzzy bounds, the predictions of $3$-separability by SSL using only the augmented states still coincide with the known results well. Our approach can be directly used to detect general multipartite entanglement, may be applied similarly to detect other multi-classification problems in quantum information theory  and the approach using partial information as the feature vectors may be used in experiment to reduce the time complexity.
\medskip

\noindent{\bf ACKNOWLEDGEMENTS}\, \, This work is supported by the National Natural Science Foundation of China (NSFC) under Grants 11571313, 12071179, 12075159 and 12171044;
Beijing Natural Science Foundation (Grant No. Z190005); the Academician Innovation Platform of Hainan Province.

\begin{appendix}

\

\

\section{The accuracy or the loss  and the epochs}

\begin{figure}[H]
\centering
\includegraphics[width=8cm]{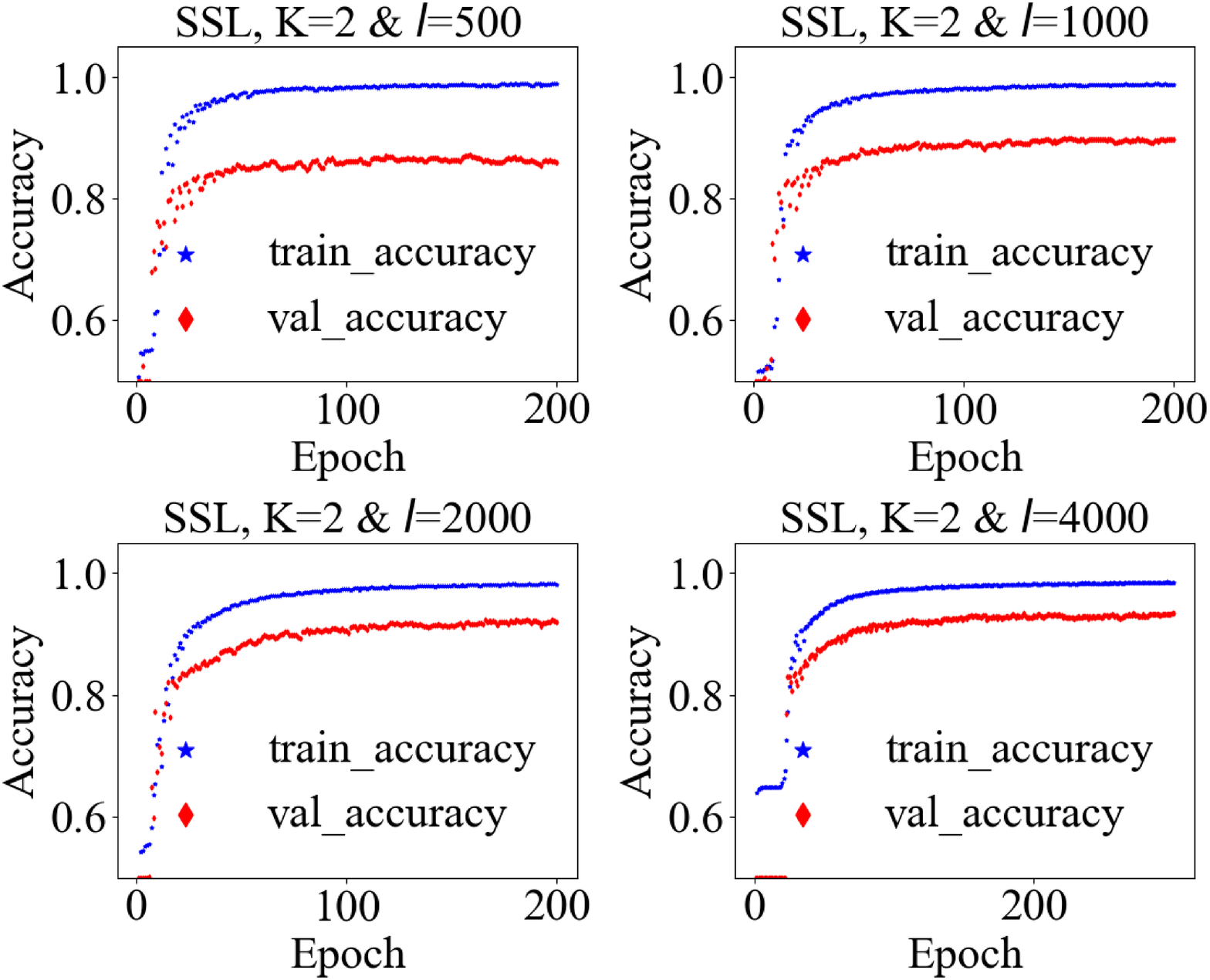}
\caption{The relationship between the epochs and the accuracy of the semi-supervised models for the entanglement detection problem of the 2-qubit quantum states when $K=2$. }
\label{epoch-acc}
\end{figure}

\begin{figure}[H]
\centering
\includegraphics[width=8cm]{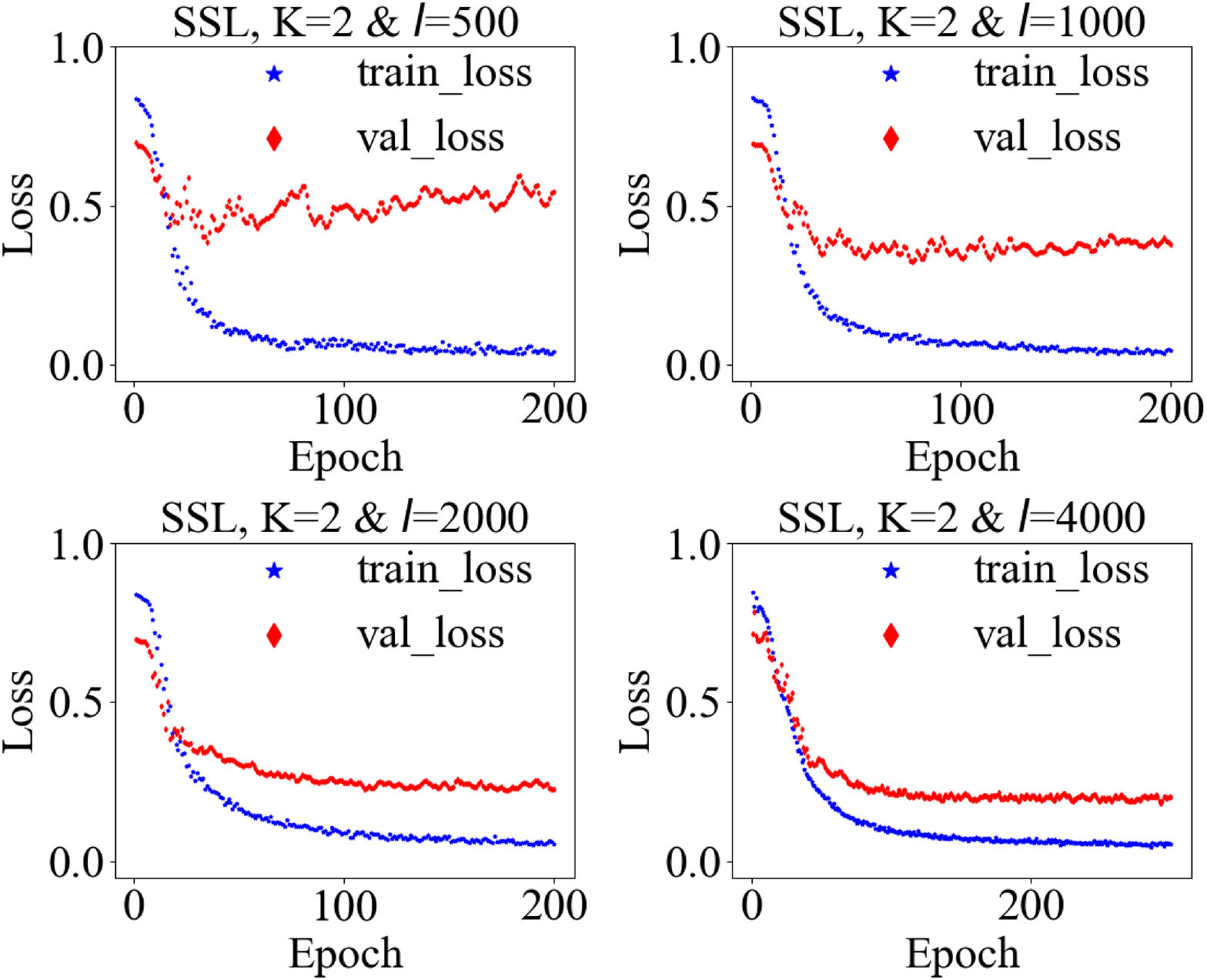}
\caption{The relationship between the epochs and the loss of the semi-supervised models for the entanglement detection problem of the 2-qubit quantum states when $K=2$.}
\label{epoch-loss}
\end{figure}

The accuracy (loss) of the training and the validation sets with respect to the epochs of the semi-supervised machine learning is shown in FIGs.\ref{epoch-acc} and \ref{epoch-loss}.
When $K=2$ and $u=20l,$ the accuracy and loss of the validation samples converge when the epochs are about 100, 100, 150 and 150 for $l=500, 1000, 2000, 4000$, respectively. Hence, the training epochs of the neural network are fixed to be 100, 100, 150 and 150, respectively.

It is worth noting that the accuracy (blue dotted line in FIG.\ref{epoch-acc}) and the loss value (blue dotted line in FIG.\ref{epoch-loss}) of the training set are calculated by using labeled data and unlabeled data with pseudo-labels. Therefore, although the SSL models reach convergence, there are always certain gaps of the accuracy and the loss between the training and validation sets as the epochs increase.

Similar to the case of $K=2$, when $K=4$ and $u = 20l$, the epochs of the SSL models of 2-qubit quantum states are fixed to be 150, 150, 200 and 250 for $l=500, 1000, 2000$ and 4000, respectively. For the entanglement detection problem of n-qubit noisy GHZ states, the epochs are fixed to be 100 empirically due to the simple structures of the noisy GHZ states.
\end{appendix}
\end{document}